\documentclass[12pt,a4paper]{article}
\usepackage[english,russian]{babel}
\usepackage{amssymb,amsmath,pifont}
\fontsize{12pt}{24pt}

\newtheorem{lem}{Lemma}
\newtheorem{teo}{Theorem}
\newtheorem{opr}{Definition}

\begin{document}

\begin{flushleft} { \bf LAPLACE INVARIANTS OF TODA LATTICES WITH THE EXCEPTIONAL

 CARTAN
MATRICES}\end{flushleft}

{\bf A.M. Guryeva\footnote[1]{Ufa State Aviation Technology
University, Ufa, Russia.} and A.V. Zhiber\footnote[2]{Institute of
Mathematics, Ufa Science Center, RAS, Ufa, Russia, e-mail:
zhiber@imat.rb.ru.}}

\begin{quote}
\hspace*{\parindent} {\small We show that Toda lattices with the
exceptional Cartan matrices  $\mathcal{G}_{2},$ $ \mathcal{F}_{4}$
and $\mathcal{E}_{6} - \mathcal{E}_{8}$ are Liouville type
systems. For these systems of equations, we obtain explicit
formulas for the invariants and generalized Laplace invariants. }
\end{quote}


\begin{flushleft}
{\bf Introduction}
\end{flushleft}

It is well known that Laplace invariants and transformations of
the scalar hyperbolic equations  were the  object of classical
research. Recently the interest to transformations and Laplace
invariants was created.  It is explained by the existence of a
close relation, discovered in a number of works \linebreak
(see~[1, 2, 6, 10-14]), between some important properties of
nonlinear equations (such as exact \linebreak integrability and
the existence of higher symmetries, conservation laws, and
differential \linebreak substitutions) and the properties of the
chain of invariants of the  linear equation. One of the most
famous examples of explicitly integrable equations in partial
derivatives is the Liouville equation
\begin{equation}
\label{math/f1}
 u_{xy}=e^{u}.
\end{equation}
There are different   notions of integrability of nonlinear
hyperbolic equations. In particular, the \linebreak definition of
the class of exactly integrable nonlinear hyperbolic equations  in
[3, 10, 13] was based on the finiteness of the chain of Laplace
invariants of the linearized equation (this equations called
equations of the Liouville type).

The natural next step in such investigations is to generalize
these results to systems of equations. In [14], the definition of
equations of the Liouville type was extended to nonlinear
hyperbolic systems of form
\begin{equation}
\label{math/f2}
 u^{i}_{xy}=F^{i}(x,y,u,u_{x},u_{y})
\end{equation}
(hereafter, we assume $u$ and $F$ to be $n$-dimensional vectors).

For completeness of the presentation in what follows, we recall
the main results in [14] pertaining to systems of differential
equations. The case of systems of differential equations involves
a major problem related to the definition of Laplace invariants.
The linearized systems of equations for systems (\ref{math/f2})
has the form
\begin{equation}
\label{math/f3} \left(D \overline{D}+aD+b \overline{D}+c
\right)v=0,
\end{equation}
where $D$ and $\overline{D}$ are the respective total derivative
operators with respect to $x$ and $y$ and where $a, b$ and $c$ are
the matrices
$$ a=-\left(\frac{\partial F^{i}}{\partial u^{j}_{x}}\right),
\qquad b=-\left(\frac{\partial F^{i}}{\partial u^{j}_{y}}\right),
\qquad c=-\left(\frac{\partial F^{i}}{\partial u^{j}}\right).$$ A
straightforward generalization of the notion of invariants to the
matrix case is as follows. \linebreak The principal Laplace
invariants are defined by the formulas
\begin{equation}
\label{math/f4}  H_{1}=D(a)+ba-c \qquad  and
 \qquad  K_{1}=\overline{D}(b)+ab-c,
\end{equation}
and the matrices $ H_{i} $  with $ i>1 $ are to be found
consecutively from the system of equations
\begin{equation}
\label{math/f5} \overline{D}( H_{i})+a_{i}H_{i}-H_{i}a_{i-1}=0,
\hspace{4.2cm}
\end{equation}
\begin{equation}
\label{math/f6}  H_{i+1}=D(a_{i})+[b,a_{i}]-
\overline{D}(b)+H_{i}, \qquad i=1,2, \ldots ,
\end{equation}
where $a_{0}=a$. If  $ H_{i} $ for $ i \leq m $ and $ a_{i} $ for
$ i \leq m-1 $ are already known, then $ a_{m} $ is determined
from Eq.~(\ref{math/f5}), and  $H_{m+1} $ is determined from Eq.
(\ref{math/f6}). But if  $ \det H_{m}=0, $ then $ a_{m} $ either
does not exist in general or is determined up to the kernels the
matrices  $ H_{m}. $ The choice of the element in the kernel
essentially affects the existence of and the explicit formulas for
the subsequent invariants.

The elements $K_{i}$ are determined similarly,
\begin{equation}
\label{math/f7} D( K_{i})+b_{i}K_{i}-K_{i}b_{i-1}=0,
\hspace{4.3cm}
\end{equation}
\begin{equation}
\label{math/f8}  K_{i+1}=\overline{D}(b_{i})+[a,b_{i}]-D(a)+K_{i},
\qquad i=1,2, \ldots ,
\end{equation}
where $b_{0}=b.$  We thus face the problem of consistently
defining the chain of invariants.

We note that the interesting systems (\ref{math/f2}) are precisely
those with degenerate matrices $H_{i}$ and $K_{i}.$

We define the elements $H_{i}$ and  $K_{i}$ somewhat differently:
if the matrices $ H_{1},H_{2}, \ldots ,H_{m}$ are known and the
equation
\begin{equation}
\label{math/f9} \overline{D} (X_{m})+a_{m}X_{m}-X_{m}a=0, \qquad
X_{m}=H_{m} \cdot H_{m-1} \cdots H_{1}, \hspace{0.1cm}
\end{equation}
has a solution $a_{m},$ we set
\begin{equation}
\label{math/f10}  H_{m+1}=D(a_{m})+[b,a_{m}]-
\overline{D}(b)+H_{m}, \quad \quad m=1,2, \ldots . \hspace{0.3cm}
\end{equation}
Similarly, if the elements $ K_{1},K_{2}, \ldots ,K_{m}$ are
already found and there exists a solution $ b_{m} $ of the
equation
\begin{equation}
\label{math/f11} D(Y_{m})+b_{m}Y_{m}-Y_{m}b=0, \qquad
Y_{m}=K_{m}\cdot K_{m-1} \cdots K_{1}, \hspace{0.4cm}
\end{equation}
 we define  $K_{m+1}$ as
\begin{equation}
\label{math/f12} K_{m+1}=\overline{D}(b_{m})+[a,b_{m}]- D(
a)+K_{m}, \quad \quad m=1,2 \ldots \;\; .
\end{equation}

It is clear that under the solvability condition for Eqs.
(\ref{math/f9}) and (\ref{math/f11}), formulas (\ref{math/f4}) and
(\ref{math/f9}) -- (\ref{math/f12}) determine a sequence of
matrices $ H_{i} $ and $K_{i},$   $ i=1,2, \ldots \; .$

We note that if relations (\ref{math/f5}) are satisfied for $
i=1,2, \ldots k$, then Eqs. (\ref{math/f9})  hold for $m=1,2,
\ldots, k $. The converse statement is not true in general.

By analogy with the scalar case, we call the matrices $H_{i}$ and
$K_{i}$ defined by formulas  (\ref{math/f4}) and \linebreak
(\ref{math/f9}) -- (\ref{math/f12}) the Laplace invariants and
also call  $X_{i}$ and  $Y_{i}, \; i=1,2, \ldots ,$ the
generalized invariants of linearized systems of equations
(\ref{math/f3}).

The conditions for the existence of solutions  $a_{m}$ and $b_{m}$
of systems of equations   (\ref{math/f9}) and  (\ref{math/f11})
are given in the following lemma.

\begin{lem}
Systems of equations (\ref{math/f9}) has a solution if and only if
the condition
\begin{equation}
\label{math/f13} \left(\overline{D}+a \right)\left(Ker X_{m}
\right) \subset Ker X_{m}
 \end{equation}
is satisfied; systems (\ref{math/f11}) has a solution if and only
if the condition
\begin{equation}
\label{math/f14} \left(D+b \right)\left(Ker Y_{m} \right) \subset
Ker Y_{m}
 \end{equation} is satisfied.
\end{lem}

 Because the matrices $a_{i}$ and $b_{i}$ determined ambiguously,
 the invariants $H_{m}$ and $K_{m}$ depend on the choice of the matrices $a_{1},a_{2}, \ldots
 ,a_{m-1}$ and  $b_{1},b_{2}, \ldots ,b_{m-1};$  therefore, the generalized Laplace invariants  $X_{m}$ and
 $Y_{m}$ are generally affected by this choice. There thus arises the question under what conditions the sequences
  $\{X_{i}\}$ and  $\{Y_{i}\}$ are well-defined. The following
  theorem answers this question.

 \begin{teo}
If
\begin{equation}
\label{math/f15}
\begin{array}{c}
  \left(D-b^{T} \right)\left(Coker
X_{i} \right) \subset Coker X_{i}, \qquad i=1,2, \ldots ,m,  \\
Coker X_{1}\subset Coker X_{2} \subset \cdots \subset Coker X_{m},
\end{array}
\end{equation}
then the generalized invariant $X_{m+1}$ is independent of the
choice of the matrices   $a_{1},a_{2}, \ldots ,a_{m}.$

 If
 \begin{equation}
\label{math/f16}
\begin{array}{c}
  \left(\overline{D}-a^{T} \right)\left(Coker
Y_{i} \right) \subset Coker Y_{i}, \qquad i=1,2, \ldots ,m,  \\
Coker Y_{1}\subset Coker Y_{2} \subset \cdots \subset Coker Y_{m},
\end{array}
\end{equation}
then the generalized invariant $Y_{m+1}$ is independent of the
choice of the matrices   $b_{1},b_{2}, \ldots ,b_{m}.$
\end{teo}

Therefore, precisely the sequences  $\{X_{i}\}$ and $\{Y_{i}\}$
(rather than $\{H_{i}\}$ and $\{K_{i}\}$) are well-defined; their
termination is placed in the basic of the definition of systems
 (\ref{math/f2}) of the Liouville type.

\begin{opr}
A system of equations (\ref{math/f2}) is said  to be a system of
the Liouville type if \linebreak conditions~(\ref{math/f13})~--~
(\ref{math/f16}) are satisfied and there exist  $\;r \geq 1 \,$
and $\; s \geq 1 \,$ such that  $ X_{r}=Y_{s}=0$.
\end{opr}

Directly generalizing Liouville  equations (\ref{math/f1}) to
systems (\ref{math/f2}) gives the truncated Toda lattices $[7, 8]$
related to the Cartan matrices of simple Lie algebras. One of the
equivalent forms of writing these systems is
\begin{equation}
\label{math/f25} u^{i}_{xy}=\sum_{j=1}^{n}a_{ij}\exp(u^{j}),
\qquad i=1,2, \ldots ,n.
\end{equation}
It is well known that these systems have $\;x$ and $\;y$ integrals
(see, e.g., [8, 9]).  We also note that, as proved in [8], systems
of equations  (\ref{math/f25}) has a complete set of $\;x$ and
$\;y$  integral if and only if the matrix $A=\left(a_{ij}\right)$
is equivalent to one of the Cartan matrices of a simple Lie
algebra.

 In this work, we show that Toda lattices  (\ref{math/f25}) with the  exceptional Cartan matrices
 $\mathcal{G}_{2},$ $ \mathcal{F}_{4},$ $ \mathcal{E}_{6}-\mathcal{E}_{8}$ are Liouville type systems
 in the sense of the definition given above.
 We obtain explicit  formulas for the invariants and generalized Laplace invariants for these systems of equations.

 We note that, as showed in [5], systems
of equations  (\ref{math/f25}) with Cartan matrices
$\mathcal{A}_{n}$, $\mathcal{B}_{n}$, $\mathcal{C}_{n},$ and
$\mathcal{D}_{n}$ are Liouville type systems.

V.V. Sokolov hypothesized that the values of the index $k$  at
which the rank of the generalized Laplace invariants  $X_{k}$
drops coincide with the exponents of the corresponding simple Lie
algebra and that the number $h$ for which  $X_{h}=0,$ is equal to
the Coxeter number.

\begin{flushleft}
{\bf 1. The Cartan matrix ${\mathcal G}_{2}$ }
\end{flushleft}

Systems of equations (\ref{math/f25}) with the Cartan matrix
${\mathcal G}_{2}$ has the form
\begin{equation}
\label{n1} u^{1}_{xy}=2\exp(u^{1})-\exp(u^{2}), \qquad
u^{2}_{xy}=-3\exp(u^{1})+2\exp(u^{2}).
\end{equation}

The invariants $H_{m}$ and generalized invariants  $X_{m}$ of the
linearized system (\ref{n1}), determined from formulas
(\ref{math/f4}), (\ref{math/f9}), and (\ref{math/f10}), are
evaluated as
$$
\begin{array}{l}
 X_{1}=H_{1}={\mathcal G}_{2}S_{1}, \qquad \quad X_{m}={\mathcal G}_{2}P^{-1}S_{m}Q, \quad
m=2,3,4,5, \qquad X_{6}=0, \medskip \\
H_{2}= {\mathcal G}_{2}P^{-1}Z_{2}{\mathcal G}_{2}^{-1}, \qquad
H_{m}=\left({\mathcal G}_{2}P^{-1}Z_{m}+Q_{m}\right)P{\mathcal
G}_{2}^{-1},
\quad m=3,4,5,6,  \medskip \\
 Q_{m}=Q_{m-1}+D(B_{m-1})+ {\mathcal G}_{2}P^{-1}D_{m}, \;\;
m=3,4,5,6, \qquad Q_{2}=0,
\end{array}
$$
where
$$ P=\left( \begin{array}{cc}
  3 & 1 \bigskip \\
  0 & \displaystyle\frac{1}{3}
\end{array} \right), \quad
Q=\left( \begin{array}{cc}
  \;\;\;1 & 0 \\
  -1 & 1
\end{array} \right), \quad
Z_{2}=\left( \begin{array}{cc}
  \;\;0 & 0 \\
  -\exp(u^{2}) & \exp(u^{1})
\end{array} \right), \quad
B_{m}=\left( \begin{array}{cc}
  b^{m}_{11} & 0 \\
   b^{m}_{21} & 0
\end{array} \right), $$ $
  b^{m}_{11}$ and $ b^{m}_{21},$   $m=2,3,4,5,
 $ are arbitrary elements, and $ D_{3}=\left( \begin{array}{cc}
   0 & 0 \\
   -\frac{1}{3}e^{u^{2}} &  0
 \end{array}\right), \; D_{m}=0, \;\;
 m=4,5,6,$ and
$$ \begin{array}{ll}
S_{1}=\mbox{diag}(\exp(u^{1}),\exp(u^{2})), \qquad \qquad
 & S_{2}=\mbox{diag}(0,\exp(u^{1}+u^{2})), \medskip  \\
S_{3}=\mbox{diag}(0,4\exp(2u^{1}+u^{2})),  &
S_{4}=\mbox{diag}(0,12\exp(3u^{1}+u^{2})), \medskip  \\
S_{5}=\mbox{diag}(0,12\exp(3u^{1}+2u^{2})),
   &  Z_{3}=\mbox{diag}(0,4\exp(u^{1})), \medskip \\
 Z_{4}=\mbox{diag}(0,3\exp(u^{1})),  & Z_{5}=\mbox{diag}(0,\exp(u^{2})), \quad Z_{6}=0.
\end{array}$$

Solutions $a_{m}$ of Eqs. (\ref{math/f9}) are then given by
\begin{equation*}
a_{1}=-{\mathcal G}_{2}R_{1}{\mathcal G}_{2}^{-1},\quad
 a_{m}=\left[
-{\mathcal G}_{2}P^{-1}R_{m}+B_{m}\right]P{\mathcal G}_{2}^{-1},
\quad m=2,3,4,5,
\end{equation*}
where $$ \begin{array}{lll}
 R_{1}=\mbox{diag}(u^{1}_{y}, u^{2}_{y}),  &
 \;\; R_{2}=\mbox{diag}(0,u^{1}_{y}+u^{2}_{y}),  &
R_{3}=\mbox{diag}(0,2u^{1}_{y}+u^{2}_{y}), \medskip \\
R_{4}=\mbox{diag}(0,3u^{1}_{y}+u^{2}_{y}),  &  \;\;
R_{5}=\mbox{diag}(0,3u^{1}_{y}+2u^{2}_{y}). &
\end{array}
$$

 We note that the values of the index $m$ at which the rank of the generalized invariants
  $X_{m}$ drops coincide with the exponents 1, 5 of the  ${\mathcal
G}_{2}$ systems, and the number  $m=6$ for which $X_{m}=0$ is
equal to the Coxeter number [4].

\begin{flushleft}
{\bf 2. The Cartan matrix ${\mathcal F}_{4}$ }
\end{flushleft}

Systems of equations (\ref{math/f25}) with the Cartan matrix
${\mathcal F}_{4}$ has the form
\begin{equation} \label{n2}
\begin{array}{l}
u^{1}_{xy}=2\exp(u^{1})-\exp(u^{2}), \\
u^{2}_{xy}=-\exp(u^{1})+2\exp(u^{2})-\exp(u^{3}), \\
u^{3}_{xy}=-2\exp(u^{2})+2\exp(u^{3})-\exp(u^{4}), \\
u^{4}_{xy}=-\exp(u^{3})+2\exp(u^{4}).
\end{array}
\end{equation}
We write this systems in the matrix form
$$ D\overline{D}u={\mathcal F}_{4}Uc, $$
where $ u=(u^{1},u^{2},u^{3},u^{4})^{T}$ is the column of the
unknowns,  $ \; c=(1,1,1,1)^{T},$  and   $
U=\mbox{diag}(\exp(u^{1}),
\\ \exp(u^{2}),\exp(u^{3}),\exp(u^{4})).$ The linearization of
Eqs.  (\ref{n2}) is then
\begin{equation}
\label{n3} D\overline{D}v={\mathcal F}_{4}Uv, \qquad
v=(v^{1},v^{2},v^{3},v^{4})^{T}.
\end{equation}

To describe the Laplace invariants and the generalized invariants,
we introduce matrices.

The matrix $J_{k}$ is the upper-triangular matrix of size $k$ all
of whose elements in the main diagonal and above it are equal to
unity. The matrix $E_{k}$ is the unit matrix of size   $k$.
\vspace{0.5cm}

The matrix $ M_{k}$ is block matrix:  $\;\;M_{k}=\left(
\begin{array}{cc}
  E_{5-k} & 0 \bigskip \\
  0 & J_{k-1}
\end{array}\right), \;\; k=2,3,4.$

We set
 $ \;\; \widetilde{J}=\left( \begin{array}{cccc}
  2 & 2 & 1 & 1 \\
  0 & 2 & 1 & 1 \\
  0 & 0 & 1 & 1 \\
  0 & 0 & 0 & 1
\end{array} \right).$
\vspace{0.5cm}

 The matrices $B_{m}, \;\; m=1,2, \ldots ,11\;$ of
size 4, are defined as

-- the first column of the matrices $ B_{m}, \; m\leq 5 $ consist
of arbitrary elements,

-- the first and second columns of the matrices $ B_{m}, \; m = 6,
7 $ consist of arbitrary elements,

 -- for $m \geq 8$ in additional to the first and second columns, the third column $ B_{m}$
 is also arbitrary,

 with the other element equal to zero.

 The diagonal matrices  $S_{m}$ and  $R_{m}$ are given by
 $$
\begin{array}{l}
  S_{1}=\mbox{diag}\left(\frac{1}{2}\exp(u^{1}), \frac{1}{2}\exp(u^{2}), \exp(u^{3}),
  \exp(u^{4})\right), \medskip
  \\
   S_{2}=\mbox{diag}\left(0, 2\exp(u^{1}+u^{2}), 2\exp(u^{2}+u^{3}), \exp(u^{3}+u^{4})\right), \medskip \\
 S_{3}=\mbox{diag}\left(0, 4\exp(2u^{2}+u^{3}), 2\exp(u^{1}+u^{2}+u^{3}),
 2\exp(u^{2}+u^{3}+u^{4})\right), \medskip \\
 S_{4}=\mbox{diag}\left(0, 18\exp(u^{1}+2u^{2}+u^{3}),
 4\exp(2u^{2}+u^{3}+u^{4}),
 2\exp(u^{1}+u^{2}+u^{3}+u^{4})\right), \medskip
 \\
  S_{5}=\mbox{diag}\left(0, 36\exp(2u^{1}+2u^{2}+u^{3}), 18\exp(u^{1}+2u^{2}+u^{3}+u^{4}),
  4\exp(2u^{2}+2u^{3}+u^{4})\right),\medskip\\
   S_{6}=\mbox{diag}\left(0, 0, 36\exp(2u^{1}+2u^{2}+u^{3}+u^{4}),
   18\exp(u^{1}+2u^{2}+2u^{3}+u^{4})\right), \medskip\\
   S_{7}=\mbox{diag}\left(0, 0, 36\exp(2u^{1}+2u^{2}+2u^{3}+u^{4}),
   18\exp(u^{1}+3u^{2}+2u^{3}+u^{4})\right), \medskip\\
   S_{8}=\mbox{diag}\left(0, 0, 0,
   18\exp(2u^{1}+3u^{2}+2u^{3}+u^{4})\right),\medskip
   \\
S_{9}=\mbox{diag}\left(0, 0, 0, 36\exp(2u^{1}+4u^{2}+2u^{3}+u^{4})\right),  \medskip\\
S_{10}=\mbox{diag}\left(0, 0, 0,
36\exp(2u^{1}+4u^{2}+3u^{3}+u^{4})\right), \medskip
\\
S_{11}=\mbox{diag}\left(0, 0, 0,
36\exp(2u^{1}+4u^{2}+3u^{3}+2u^{4})\right),
\quad S_{12}=0, \medskip\\
 R_{1}=\mbox{diag}\left(u^{1}_{y},
u^{2}_{y}, u^{3}_{y}, u^{4}_{y}\right), \medskip
  \\
   R_{2}=\mbox{diag}\left(0, u^{1}_{y}+u^{2}_{y}, u^{2}_{y}+u^{3}_{y}, u^{3}_{y}+u^{4}_{y}\right), \medskip \\
 R_{3}=\mbox{diag}\left(0, 2u^{2}_{y}+u^{3}_{y}, u^{1}_{y}+u^{2}_{y}+u^{3}_{y},
 u^{2}_{y}+u^{3}_{y}+u^{4}_{y}\right), \medskip \\
 R_{4}=\mbox{diag}\left(0, u^{1}_{y}+2u^{2}_{y}+u^{3}_{y}, 2u^{2}_{y}+u^{3}_{y}+u^{4}_{y},
 u^{1}_{y}+u^{2}_{y}+u^{3}_{y}+u^{4}_{y}\right), \medskip \\
  R_{5}=\mbox{diag}\left(0, 2u^{1}_{y}+2u^{2}_{y}+u^{3}_{y}, u^{1}_{y}+2u^{2}_{y}+u^{3}_{y}+u^{4}_{y},
  2u^{2}_{y}+2u^{3}_{y}+u^{4}_{y}\right), \medskip \\
   R_{6}=\mbox{diag}\left(0, 0, 2u^{1}_{y}+2u^{2}_{y}+u^{3}_{y}+u^{4}_{y},
   u^{1}_{y}+2u^{2}_{y}+2u^{3}_{y}+u^{4}_{y}\right), \medskip \\
   R_{7}=\mbox{diag}\left(0, 0, 2u^{1}_{y}+2u^{2}_{y}+2u^{3}_{y}+u^{4}_{y},
   u^{1}_{y}+3u^{2}_{y}+2u^{3}_{y}+u^{4}_{y}\right), \medskip \\
   R_{8}=\mbox{diag}\left(0, 0, 0, 2u^{1}_{y}+3u^{2}_{y}+2u^{3}_{y}+u^{4}_{y}\right),
  \medskip  \\
R_{9}=\mbox{diag}\left(0, 0, 0, 2u^{1}_{y}+4u^{2}_{y}+2u^{3}_{y}+u^{4}_{y}\right), \medskip \\
R_{10}=\mbox{diag}\left(0, 0, 0,
2u^{1}_{y}+4u^{2}_{y}+3u^{3}_{y}+u^{4}_{y}\right), \medskip
\\
R_{11}=\mbox{diag}\left(0, 0, 0,
2u^{1}_{y}+4u^{2}_{y}+3u^{3}_{y}+2u^{4}_{y}\right).
\end{array}
$$

The matrices  $Z_{m}, \;\; m=1,2, \ldots ,12,$ are given by  $
\;\;Z_{1}=U ,$
\begin{equation*}
\begin{array}{ll}
Z_{2}=\left(\begin{array}{cccc}
  0 & 0 & 0 & 0 \\
  -2e^{u^{2}} & 2e^{u^{1}} & 0 & 0 \\
  0 & -2e^{u^{3}} & 2e^{u^{2}} & 0 \\
  0 & 0 & -e^{u^{4}} & e^{u^{3}}
\end{array} \right), & \quad Z_{3}=\left(\begin{array}{cccc}
  0 & 0 & 0 & 0 \\
  0 & 0 & 2e^{u^{2}} & 0 \\
  0 & -e^{u^{3}} & e^{u^{1}} & 0 \\
  0 & 0 & -e^{u^{4}} & 2e^{u^{2}}
  \end{array} \right),  \bigskip \\
Z_{4}=\left(\begin{array}{cccc}
  0 & 0 & 0 & 0 \\
  0 & 3e^{u^{1}} & 3e^{u^{2}} & 0 \\
  0 & -e^{u^{4}} & 0 & 2e^{u^{2}}  \\
  0 & 0 & -e^{u^{4}} & e^{u^{1}}
\end{array} \right), &  \quad Z_{5}=\left(\begin{array}{cccc}
  0 & 0 & 0 & 0 \\
  0 & 2e^{u^{1}} & 0 & 0 \\
  0 & -e^{u^{4}} & 3e^{u^{1}} & 3e^{u^{2}} \\
  0 & 0 & -e^{u^{3}} & 0
\end{array} \right),
\end{array}
\end{equation*}
\begin{equation*}
\begin{array}{ll}
Z_{6}=\left(\begin{array}{cccc}
  0 & 0 & 0 & 0 \\
  0 & 0 & 0 & 0 \\
  0 & -e^{u^{4}} & 2e^{u^{1}} & 0  \\
  0 & 0 & -e^{u^{3}} & 3e^{u^{1}}
\end{array} \right), &  \quad
Z_{7}=\left(\begin{array}{cccc}
  0 & 0 & 0 & 0 \\
  0 & 0 & 0 & 0 \\
  0 & 0 & e^{u^{3}} & -2e^{u^{1}}  \\
  0 & 0 & 0 & e^{u^{2}}
\end{array} \right),
 \\
Z_{8}=\left(\begin{array}{cccc}
  0 & 0 & 0 & 0 \\
  0 & 0 & 0 & 0 \\
  0 & 0 & 0 & 0  \\
  0 & 0 & -e^{u^{2}} & e^{u^{1}}
\end{array} \right), &  \quad
Z_{9}=\left(\begin{array}{cccc}
  0 & 0 & 0 & 0 \\
  0 & 0 & 0 & 0 \\
  0 & 0 & 0 & 0  \\
  0 & 0 & -e^{u^{2}} & 2e^{u^{2}}
\end{array} \right),
\\
Z_{10}=\left(\begin{array}{cccc}
  0 & 0 & 0 & 0 \\
  0 & 0 & 0 & 0 \\
  0 & 0 & 0 & 0  \\
  0 & 0 & -e^{u^{2}} & e^{u^{3}}
\end{array} \right), &  \quad
Z_{11}=\left(\begin{array}{cccc}
  0 & 0 & 0 & 0 \\
  0 & 0 & 0 & 0 \\
  0 & 0 & 0 & 0  \\
  0 & 0 & -e^{u^{2}} & e^{u^{4}}
\end{array} \right),
\\
Z_{12}=\left(\begin{array}{cccc}
  0 & 0 & 0 & 0 \\
  0 & 0 & 0 & 0 \\
  0 & 0 & 0 & 0  \\
  0 & 0 & -e^{u^{2}} & 0
\end{array} \right). &  \qquad
\end{array}
\end{equation*}

The matrices $P_{m}$ and $D_{m}, \;\; m=2, 3, \ldots , 12,$ are
given by
$$
P_{2}=E_{4}, \quad    P_{3}=\left(\begin{array}{cccc}
  1 & 0 & 0 & 0 \\
  0 & 1 & 0 & 0 \\
  0 & -1 & 1 & 0  \\
  0 & 0 & 0 & 1
\end{array} \right), \quad  P_{4}=\left(\begin{array}{cccc}
  1 & 0 & 0 & 0 \\
  0 & 1 & 0 & 0 \\
  0 & -1/3 & 1 & 0  \\
  0 & 1/3 & -1 & 1
\end{array} \right), $$

$$ \begin{array}{lll} P_{5}=\left(\begin{array}{cccc}
  1 & 0 & 0 & 0 \\
  0 & 1 & -1 & 3 \\
  0 & 0 & 0 & 3  \\
  0 & 0 & -1 & 3
\end{array} \right),  \quad &

P_{6}=\left(\begin{array}{cccc}
  1 & 0 & 0 & 0 \\
  0 & 1 & 0 & \frac{1}{2} \\
  0 & -1 & 1 & 0  \\
  0 & 0 & 0 & 3/2
\end{array} \right), \quad & P_{7}=E_{4},
\bigskip \\
P_{8}=\left(\begin{array}{cccc}
  1 & 0 & 0 & 0 \\
  0 & 1 & 0 & 0 \\
  0 & 0 & 1 & 2  \\
  0 & 0 & 0 & 1
\end{array} \right),  \quad &  P_{9}=P_{10}=P_{11}=P_{12}=E_{4}, &  \bigskip
\end{array} $$

$$
 \begin{array}{lll}
D_{3}=\left(\begin{array}{cccc}
  0 & 0 & 0 & 0 \\
  -e^{u^{2}} & 0 & 0 & 0 \\
  0 & 0 & 0 & 0  \\
  0 & 0 & 0 & 0
\end{array} \right), \quad & D_{7}=\left(\begin{array}{cccc}
  0 & 0 & 0 & 0 \\
  0 & 0 & 0 & 0 \\
  0 & 2e^{u^{1}} & 0 & 0  \\
  0 & -e^{u^{3}} & 0 & 0
\end{array} \right), \bigskip \\

 D_{k}=0, \quad k=2,4,5,6,8,9, \ldots ,12. &
\end{array}
$$

We have the following theorem.
\begin{teo} Systems of equations  (\ref{n2}) is a Liouville type system. The generalized invariants
$X_{m},$ the  invariants $H_{m}$ of linearized system (\ref{n3}),
determined from formulas (\ref{math/f4}), (\ref{math/f9}) and
(\ref{math/f10}), are evaluated as
 \begin{equation*}
\begin{array}{l}
  X_{m}={\mathcal F}_{4}\widetilde{J}^{-1}A_{m}S_{m}A_{m}^{T}\left(J^{-1}_{4} \right)^{T}, \quad m=1,2, \ldots , 11,
 \quad X_{12}=0, \medskip  \\
   H_{m}=\left[ {\mathcal F}_{4}\widetilde{J}^{-1}A_{m}Z_{m}+Q_{m-1}\right ]A_{m-1}^{-1}
  \widetilde{J}{\mathcal F}_{4}^{-1}, \;\; \qquad m=1,2, \ldots ,12,
\end{array}
\end{equation*}
where the matrices $A_{m}$ and $Q_{m-1}$ are evaluated using the
recursive relations
\begin{equation*}
\begin{array}{l}
  A_{m}=A_{m-1}M_{k}^{-1}P_{m}^{-1}, \qquad m=3,4, \ldots ,12, \medskip \\
  k=4 \;\;\mbox{при} \;\; m\leq 6, \;\; k=3\;\;  \mbox{при}\;\; m = 7,\;\; k=2\;\;  \mbox{при}\;\; m\geq 8, \medskip \\
 A_{0}=\widetilde{J}{\mathcal F}_{4}^{-1}, \qquad
 A_{1}=\widetilde{J}, \qquad A_{2}=E_{4},  \medskip \\
  Q_{m-1}={\mathcal F}_{4}\widetilde{J}^{-1}A_{m-1}D_{m}+D(B_{m-1})+Q_{m-2}A_{m-2}^{-1}A_{m-1},
\end{array}
\end{equation*}

$m=2,3, \ldots , 12, \;\; Q_{0}=0.$

The elements  $a_{m}$ are then given by
\begin{equation*}
 a_{m}=\left[
 -{\mathcal
F}_{4}\widetilde{J}^{-1}A_{m}R_{m}+B_{m}\right]A_{m}^{-1}\widetilde{J}{\mathcal
F}_{4}^{-1}, \qquad m=1,2, \ldots ,11.
\end{equation*} \end{teo}

The matrices  $ S_{m}, $ $ R_{m}, $ and  $ Z_{m} $ were calculated
by means of  programm <<Invariant>>, intended for  Maple~V Release
4.

We note that the values of the index $m$ at which the rank of the
generalized invariants
  $X_{m}$ drops coincide with the exponents 1, 5, 7, 11 of the  ${\mathcal
F}_{4}$ systems, and the number  $m=12$ for which $X_{m}=0$ is
equal to the Coxeter number [4].

\begin{flushleft}
{\bf 3. The Cartan matrices ${\mathcal E}_{6}-{\mathcal E}_{8}$ }
\end{flushleft}

Systems of equations (\ref{math/f25}) with the Cartan matrix
${\mathcal E}_{\alpha}, \;\; \alpha=6,7,8 $   has the form
\begin{equation} \label{n4}
\begin{array}{l}
u^{1}_{xy}=2\exp(u^{1})-\exp(u^{3}), \\
u^{2}_{xy}=2\exp(u^{2})-\exp(u^{4}), \\
u^{3}_{xy}=-\exp(u^{1})+2\exp(u^{3})-\exp(u^{4}), \\
u^{4}_{xy}=-\exp(u^{2})-\exp(u^{3})+2\exp(u^{4})-\exp(u^{5}), \\
u^{i}_{xy}=-\exp(u^{i-1})+2\exp(u^{i})-\exp(u^{i+1}), \qquad \;\; i=5,6, \ldots , \alpha-1,  \\
u^{\alpha}_{xy}=-\exp(u^{\alpha-1})+2\exp(u^{\alpha}).
\end{array}
\end{equation}

The linearization of Eqs. (\ref{n4}) is then
 \begin{equation}
\label{n5} D\overline{D}v={\mathcal E}_{\alpha}Uv.
\end{equation}

The result in this section can be formulated as follows.
\begin{teo}
Systems of equations  (\ref{n4}) is a Liouville type system. The
generalized invariants  $X_{m},$ the invariants $H_{m}$ of
linearized system (\ref{n5}), determined from formulas
(\ref{math/f4}), (\ref{math/f9}), and (\ref{math/f10}), are
evaluated as
\begin{equation*}
\begin{array}{l}
  X_{m}={\mathcal E}_{\alpha}A_{m}S_{m}A_{m}^{T}, \quad m=1,2, \ldots , r-1,
 \quad X_{r}=0, \medskip \\
   H_{m}=\left[ {\mathcal E}_{\alpha}A_{m}Z_{m}+Q_{m-1}\right ]A_{m-1}^{-1}
  {\mathcal E}_{\alpha}^{-1}, \;\; \qquad m=1,2, \ldots ,r,
\end{array}
\end{equation*}
where matrices $A_{m}$ and $Q_{m-1}$ are evaluated using the
recursive relations
\begin{equation*}
\begin{array}{l}
  A_{m}=A_{m-1} G_{m}^{-1}P_{m}^{-1}, \;\;\; m=2,4, \ldots , r, \quad A_{0}= {\mathcal E}_{\alpha}^{-1}, \;\;
 A_{1}= E_{\alpha},  \medskip \\
    Q_{m-1}={\mathcal
    E}_{\alpha}A_{m-1}D_{m}+D(B_{m-1})+Q_{m-2}A_{m-2}^{-1}A_{m-1},
    \;\; m=2,3, \ldots , r, \;\; Q_{0}=0.
\end{array}
\end{equation*}

The elements $a_{m}$ are then given by
\begin{equation*}
 a_{m}=\left[
-{\mathcal E}_{\alpha}A_{m}R_{m}+B_{m}\right]A_{m}^{-1}{\mathcal
E}_{\alpha}^{-1}, \qquad m=1,2, \ldots , r-1.
\end{equation*} \end{teo}

The number $r$ is equal 12  if  $\alpha $ set to 6; r is equal 18
if   $\alpha $ set to 7 and r is equal 30 if $\alpha$ set to 8.

The matrices  $ S_{m}, $ $ R_{m}, $ and  $ Z_{m} $ were calculated
by means of  programm <<Invariant>>, intended for  Maple~V Release
4.

 The matrices  $P_{m},\;D_{m}, \;G_{m}\;
B_{m},\;S_{m},\; R_{m},\;Z_{m}$ are given by formulaes for each
$\alpha=6,7,8\;$ their.

We recall that  $J_{k}$ is the upper-triangular matrix of size $k$
all of whose elements in the main diagonal and above it are equal
to unity, and the matrix $E_{k}$ is the unit matrix of size $k$.

 To find the Laplace invariants of Eqs. (\ref{n5}) when $\alpha=6,$  we introduce matrices of size 6.

The matrices  $G_{m}, \; \; m=2,3, \ldots ,12$ are given by
$\;\;G_{2}=J_{6},$
$$
\begin{array}{l}
 G_{m}=\left(\begin{array}{cc}
   1 & 0 \\
   0 &  J_{5}
 \end{array} \right), \quad  m=3,4,5; \quad
  G_{6}=\left(\begin{array}{ccc}
   1 & 0   & 0\\
   0 &  J_{4} & 0 \\
   0 & 0 & 1
 \end{array} \right); \\
  G_{m}=\left(\begin{array}{ccc}
   1 & 0   & 0\\
   0 &  J_{3} & 0 \\
   0 & 0 & E_{2}
 \end{array} \right), \quad m=7,8; \quad G_{m}=E_{6} \;\;
for \;\; m=9,10, 11,12.
\end{array}
$$

The matrices $B_{m}, \;\; m=1,2, \ldots ,11\;$ of size 6, are
defined as

\begin{itemize}
  \item [--] the first column of the matrices $ B_{m}, \; m\leq 4 $ consist
of arbitrary elements,

\item [--] for $m =5 $ in additional to the first column, the last column $
B_{m}$ is also arbitrary,

\item [--] the first, fifth  and sixth columns of the matrices $ B_{m}, \;
m = 6, 7 $ consist of arbitrary elements,

\item [--] the first,  fourth, fifth  and sixth columns of the matrices $
B_{8} $ consist of arbitrary elements,

\item [--] for $m \geq 9 $ in additional to the first,  fourth, fifth  and
sixth columns, the third column $ B_{m}$
 is also arbitrary,
\end{itemize}

with the other element equal to zero.

\vspace{0.2cm}

 The matrices $P_{m}, \;\; m=2,3, \ldots ,12$ are
given by

$$
\begin{array}{cc}
P_{2}=\left (\begin{array}{lrrrrr}
  1 & 0 & 0 & \;\;0 & \;\;0 & \;\;0 \\
  0 & 1 & 0 & 0 & 0 & 0 \\
  0 & -1 & 1 & 0 & 0 & 0 \\
  0 & 1 & -1 & 1 & 0 & 0 \\
  0 & 0 & 0 & 0 & 1 & 0 \\
  0 & 0 & 0 & 0 & 0 & 1
\end{array} \right), \quad &  P_{3}=\left (\begin{array}{lrrrrr}
  1 & 0 & 0 & 0 & \;\;0 & \;\;0 \\
  0 & 1 & 0 & 0 & 0 & 0 \\
  0 & -1 & 1 & 0 & 0 & 0 \\
  0 &  \frac{1}{2} & -1 & 1 & 0 & 0 \\
  0 & - \frac{1}{2} & 1 & -1 & 1 & 0 \\
  0 & 0 & 0 & 0 & 0 & 1
\end{array} \right),
\end{array}
 $$
$$
\begin{array}{cc}
 P_{4}=\left (\begin{array}{lrrrrr}
  1 & 0 & 0 & 0 & \;\;0 & \;\;0 \\
  0 & 1 & 0  & 0 & 0 & 0 \\
  0 & -\frac{1}{3}  & 1  & 0 & 0 & 0 \\
  0 &  \frac{1}{3} & -1  & 1 & 0 & 0 \\
  0 & - \frac{1}{6} & \frac{1}{2}  & -1 & 1 & 0 \\
  0 & \frac{1}{6}  & -\frac{1}{2}  & 1 & -1 & 1
\end{array} \right),  \quad &

 P_{5}=\left (\begin{array}{lrrrrr}
  1 & 0 & 0 & 0 & \;\;0 & \;\;0 \\
  0 & 1 & 0  & 0 & 0 & 0 \\
  0 & -1  & 1  & 0 & 0 & 0 \\
  0 &  \frac{1}{3} & -\frac{1}{3}  & 1 & 0 & 0 \\
  0 & -\frac{1}{3} & \frac{1}{3}  & -1 & 1 & 0 \\
  0 & \frac{1}{6}  & -\frac{1}{6}  & \frac{1}{2} & -1 & 1
\end{array} \right), \bigskip \\

P_{6}=\left (\begin{array}{lrrrrr}
  1 & 0 & 0 & 0 & \;\;0 & \;\;0 \\
  0 & 1 & 0  & 0 & 0 & 0 \\
    0 &  -\frac{3}{5} & 1  & 0 & 0 & 0 \\
   0 &  \frac{3}{5} & -1  & 1 & 0 & 0 \\
  0 & -\frac{1}{5} & \frac{1}{3}  & -\frac{1}{3} & 1 & 0 \\
  0 & 0  & 0  & 0 & 0 & 1
\end{array} \right), \quad &
P_{7}=\left (\begin{array}{lrrrrr}
  1 & \;\;0 & 0 & \;\;0 & \;\;0 & \;\;0 \\
  0 & 1 & -1 & \frac{5}{3} & 0 & 0 \\
  0 & 0 & 1 & -\frac{5}{3} & 0 & 0 \\
  0 & 0 & \frac{5}{3} & 0 & 0 & 0 \\
  0 & 0 & 0 & 0 & 1 & 0 \\
  0 & 0 & 0 & 0 & 0 & 1
\end{array} \right), \bigskip
\end{array}
 $$
 $$
 \begin{array}{cc}
P_{8}=\left (\begin{array}{lrrrrr}
  1 & \;\;0 & 0 & \;\;0 & \;\;0 & \;\;0 \\
  0 &  \frac{5}{7} & 0 & 0 & 0 & 0 \\
  0 & -1 & 1 & 0 & 0 & 0 \\
  0 & \frac{2}{7} & -1 & 1 & 0 & 0 \\
  0 & 0 & 0 & 0 & 1 & 0 \\
  0 & 0 & 0 & 0 & 0 & 1
\end{array} \right), \quad &
P_{9}=\left (\begin{array}{lrrrrr}
  1 & \;\;0 & \;\;0 & \;\;0 & \;\;0 & \;\;0 \\
  0 &  1 & 0 & 0 & 0 & 0 \\
  0 & -1 & 1 & 0 & 0 & 0 \\
  0 & 0 & 0 & 1 & 0 & 0 \\
  0 & 0 & 0 & 0 & 1 & 0 \\
  0 & 0 & 0 & 0 & 0 & 1
\end{array} \right), \bigskip \\
P_{10}=P_{11}=P_{12}=E_{6}. \hspace{3cm} \\
\end{array}
 $$

 The nonzero elements of the matrices $D_{m}=\left(d_{ij}\right)_{i,j=1}^{6}, \; m=2,3, \ldots ,12,
 $ are determined by the formulas
$$
\begin{array}{l}
d^{3}_{2 1}=-e^{u^{3}}; \;\; d^{6}_{4 6}=2e^{u^{2}}-2e^{u^{3}};
\;\; d^{6}_{5 6}=e^{u^{1}}; \;\; d^{7}_{3 5}=-e^{u^{4}}; \; \;
d^{7}_{5 5}=3e^{u^{2}}-\frac{3}{2}e^{u^{1}}; \medskip \\ d^{9}_{2
4}=e^{u^{5}}-\frac{5}{2}e^{u^{1}}; \quad d^{9}_{3 5}=e^{u^{3}};
\quad d^{10}_{2 3}=e^{u^{5}}.
\end{array}
 $$

The diagonal matrices  $S_{m}$ and $R_{m},$ are given by
\begin{equation*}
\begin{array}{l}
S_{1}=\mbox{diag}\left(\exp(u^{1}), \exp(u^{2}), \exp(u^{3}),
\exp(u^{4}), \exp(u^{5}), \exp(u^{6})\right),  \medskip \\
  S_{2}=\mbox{diag}\left(0, \exp(u^{1}+u^{3}), \exp(u^{2}+u^{4}),
\exp(u^{3}+u^{4}), \exp(u^{4}+u^{5}), \exp(u^{5}+u^{6})\right),
\medskip \\  S_{3}=\mbox{diag}\left(0,
4\exp(u^{2}+u^{3}+u^{4}), \exp(u^{1}+u^{3}+u^{4}),
\exp(u^{2}+u^{4}+u^{5}), \exp(u^{3}+u^{4}+u^{5}), \right. \medskip
\\ \left.
\exp(u^{4}+u^{5}+u^{6})\right), \qquad
 S_{4}=\mbox{diag}\left(0,
9\exp(u^{1}+u^{2}+u^{3}+u^{4}), 4\exp(u^{2}+u^{3}+u^{4}+u^{5}),
\right. \medskip \\ \left. \exp(u^{1}+u^{3}+u^{4}+u^{5}),
\exp(u^{2}+u^{4}+u^{5}+u^{6}),
\exp(u^{3}+u^{4}+u^{5}+u^{6})\right), \medskip \\
 S_{5}=\mbox{diag}\left(0,
4\exp(u^{2}+u^{3}+2u^{4}+u^{5}), 9\exp(u^{1}+u^{2}+u^{3}+u^{4}+u^{5}), \right. \medskip \\
\left. 4\exp(u^{2}+u^{3}+u^{4}+u^{5}+u^{6}),
\exp(u^{1}+u^{3}+u^{4}+u^{5}+u^{6}), 0 \right), \medskip \\
S_{6}=\mbox{diag}\left(0,
25\exp(u^{1}+u^{2}+u^{3}+2u^{4}+u^{5}), 4\exp(u^{2}+u^{3}+2u^{4}+u^{5}+u^{6}), \right. \medskip \\
\left.
9\exp(u^{1}+u^{2}+u^{3}+u^{4}+u^{5}+u^{6}), 0, 0 \right), \medskip \\
S_{7}=\mbox{diag}\left(0, 25\exp(u^{1}+u^{2}+2u^{3}+2u^{4}+u^{5}),
4\exp(u^{1}+u^{2}+u^{3}+2u^{4}+\right. \medskip \\ \left.
+2u^{5}+u^{6}),
25\exp(u^{2}+u^{3}+2u^{4}+u^{5}+u^{6}), 0, 0 \right),\medskip \\
\end{array}
\end{equation*}
\begin{equation*}
\begin{array}{l}
S_{8}=\mbox{diag}\left(0,
25\exp(u^{1}+u^{2}+u^{3}+2u^{4}+2u^{5}+u^{6}),  \right.\medskip \\
 \left. 25\exp(u^{1}+u^{2}+2u^{3}+2u^{4}+u^{5}+u^{6}),  0,0, 0 \right),\medskip \\
S_{9}=\mbox{diag}\left(0,
25\exp(u^{1}+u^{2}+u^{3}+2u^{4}+2u^{5}+u^{6}),  0, 0 ,0,0\right), \medskip \\
S_{10}=\mbox{diag}\left(0,
25\exp(u^{1}+u^{2}+2u^{3}+3u^{4}+2u^{5}+u^{6}),  0, 0, 0, 0 \right), \medskip\\
S_{11}=\mbox{diag}\left(0,
25\exp(u^{1}+2u^{2}+2u^{3}+3u^{4}+2u^{5}+u^{6}),  0, 0, 0, 0
\right),
\quad  S_{12}=0,\medskip \\
 R_{1}=\mbox{diag}\left(u^{1}_{y}, u^{2}_{y}, u^{3}_{y},
u^{4}_{y}, u^{5}_{y}, u^{6}_{y} \right), \medskip\\
  R_{2}=\mbox{diag}\left(0, u^{1}_{y}+u^{3}_{y}, u^{2}_{y}+u^{4}_{y},
u^{3}_{y}+u^{4}_{y}, u^{4}_{y}+u^{5}_{y},
u^{5}_{y}+u^{6}_{y})\right), \medskip\\
 R_{3}=\mbox{diag}\left(0,
u^{2}_{y}+u^{3}_{y}+u^{4}_{y}, u^{1}_{y}+u^{3}_{y}+u^{4}_{y},
u^{2}_{y}+u^{4}_{y}+u^{5}_{y}, u^{3}_{y}+u^{4}_{y}+u^{5}_{y},
u^{4}_{y}+u^{5}_{y}+u^{6}_{y} \right), \medskip \\
R_{4}=\mbox{diag}\left(0, u^{1}_{y}+u^{2}_{y}+u^{3}_{y}+u^{4}_{y},
u^{2}_{y}+u^{3}_{y}+u^{4}_{y}+u^{5}_{y},
u^{1}_{y}+u^{3}_{y}+u^{4}_{y}+u^{5}_{y},
u^{2}_{y}+u^{4}_{y}+u^{5}_{y}+u^{6}_{y}, \right. \medskip \\
\left. u^{3}_{y}+u^{4}_{y}+u^{5}_{y}+u^{6}_{y} \right), \quad
R_{5}=\mbox{diag}\left(0,
u^{2}_{y}+u^{3}_{y}+2u^{4}_{y}+u^{5}_{y}, u^{1}_{y}+u^{2}_{y}+u^{3}_{y}+u^{4}_{y}+u^{5}_{y}, \right. \medskip \\
\left. u^{2}_{y}+u^{3}_{y}+u^{4}_{y}+u^{5}_{y}+u^{6}_{y},
u^{1}_{y}+u^{3}_{y}+u^{4}_{y}+u^{5}_{y}+u^{6}_{y}, 0 \right), \medskip \\
R_{6}=\mbox{diag}\left(0,
u^{1}_{y}+u^{2}_{y}+u^{3}_{y}+2u^{4}_{y}+u^{5}_{y}, u^{2}_{y}+u^{3}_{y}+2u^{4}_{y}+u^{5}_{y}+u^{6}_{y}, \right. \medskip\\
\left.
u^{1}_{y}+u^{2}_{y}+u^{3}_{y}+u^{4}_{y}+u^{5}_{y}+u^{6}_{y}, 0, 0 \right),\medskip \\
R_{7}=\mbox{diag}\left(0,
u^{1}_{y}+u^{2}_{y}+2u^{3}_{y}+2u^{4}_{y}+u^{5}_{y}, u^{1}_{y}+u^{2}_{y}+u^{3}_{y}+2u^{4}_{y}+2u^{5}_{y}+u^{6}_{y}, \right.
\medskip \\
\left. u^{2}_{y}+u^{3}_{y}+2u^{4}_{y}+u^{5}_{y}+u^{6}_{y}, 0, 0
\right), \medskip \\
 R_{8}=\mbox{diag}\left(0,
u^{1}_{y}+u^{2}_{y}+u^{3}_{y}+2u^{4}_{y}+2u^{5}_{y}+u^{6}_{y},
\right. \medskip \\ \left.
u^{1}_{y}+u^{2}_{y}+u^{3}_{y}+2u^{4}_{y}+
2u^{5}_{y}+u^{6}_{y},  0, 0,0 \right), \medskip \\
R_{9}=\mbox{diag}\left(0,
u^{1}_{y}+u^{2}_{y}+2u^{3}_{y}+2u^{4}_{y}+2u^{5}_{y}+u^{6}_{y},  0, 0, 0, 0 \right), \medskip \\
R_{10}=\mbox{diag}\left(0,
u^{1}_{y}+u^{2}_{y}+2u^{3}_{y}+3u^{4}_{y}+2u^{5}_{y}+u^{6}_{y}, 0,
0, 0, 0 \right), \medskip\\ R_{11}=\mbox{diag}\left(0,
u^{1}_{y}+2u^{2}_{y}+2u^{3}_{y}+3u^{4}_{y}+2u^{5}_{y}+u^{6}_{y},
0, 0, 0 ,0 \right).
\end{array}
\end{equation*}

\vspace{0.3cm}

 The matrices $Z_{m}, \;\; m=1,2, \ldots ,12,$ are
given by $ \;\;Z_{1}=U ,$
\begin{equation*}
\begin{array}{ll}
 Z_{2}=\left(\begin{array}{cccccc}
  0 & 0 & 0 & 0 & 0 & 0 \\
  -e^{u^{3}} & 0 & e^{u^{1}} & 0 & 0 & 0  \\
  0 & -e^{u^{4}} & 0 & e^{u^{2}} & 0 & 0 \\
  0 & 0 & -e^{u^{4}} & e^{u^{3}} & 0 &0 \\
   0 & 0 & 0 & -e^{u^{5}} & e^{u^{4}} & 0  \\
    0 & 0 & 0 & 0 & -e^{u^{6}} & e^{u^{5}}
\end{array} \right), \quad
Z_{3}=\left(\begin{array}{cccccc}
  0 & 0 & 0 & 0 & 0 & 0 \\
  0 & 0 & 2e^{u^{3}} & 2e^{u^{2}} & 0 & 0  \\
  0 & -e^{u^{4}} & 0 & e^{u^{1}} & 0 & 0 \\
  0 & 0 & -e^{u^{5}} & 0 & e^{u^{2}} &0 \\
   0 & 0 & 0 & -e^{u^{5}} & e^{u^{3}} & 0  \\
    0 & 0 & 0 & 0 & -e^{u^{6}} & e^{u^{4}}
\end{array} \right),
\end{array}
\end{equation*}
\begin{equation*}
\begin{array}{ll}
Z_{4}=\left(\begin{array}{cccccc}
  0 & 0 & 0 & 0 & 0 & 0 \\
  0 & \frac{3}{2}e^{u^{1}} & 3e^{u^{2}} & 0 & 0 & 0  \\
  0 & -e^{u^{5}} & 0 & 2e^{u^{3}} & 2e^{u^{2}} & 0 \\
  0 & 0 & -e^{u^{5}} & 0 & e^{u^{1}} &0 \\
   0 & 0 & 0 & -e^{u^{6}} & 0 & e^{u^{2}}   \\
    0 & 0 & 0 & 0 & -e^{u^{6}} & e^{u^{3}}
\end{array} \right), \quad
Z_{5}=\left(\begin{array}{cccccc}
  0 & 0 & 0 & 0 & 0 & 0 \\
  0 & 0 & e^{u^{4}} & 0 & 0 & 0  \\
  0 & -e^{u^{5}} & \frac{3}{2}e^{u^{1}} & 3e^{u^{2}} & 0 & 0 \\
  0 & 0 & -e^{u^{6}} & 0 & 2e^{u^{3}} & 2e^{u^{2}} \\
   0 & 0 & 0 & -e^{u^{6}} & 0 & e^{u^{1}}   \\
    0 & 0 & 0 & 0 & 0 & 0
\end{array} \right),
\end{array}
\end{equation*}
\begin{equation*}
\begin{array}{ll}
Z_{6}=\left(\begin{array}{cccccc}
  0 & 0 & 0 & 0 & 0 & \;\;0 \\
  0 & \frac{5}{2}e^{u^{1}} & \frac{5}{3}e^{u^{4}} & 0 & 0 & \;\;0  \\
  0 & -e^{u^{6}} & 0 & e^{u^{4}} & 0 & \;\;0 \\
  0 & 0 & -e^{u^{6}} & \frac{3}{2}e^{u^{1}} & 3e^{u^{2}} & \;\;0 \\
   0 & 0 & 0 & 0 & 0 & \;\; 0   \\
    0 & 0 & 0 & 0 & 0 & \;\;0
\end{array} \right), &
Z_{7}=\left(\begin{array}{cccccc}
  0 & 0 & 0 & 0 &\;\; 0 & \;\;0 \\
  0 & e^{u^{3}} & 0 & 0 & \;\;0 & \;\;0  \\
  0 & 0 & e^{u^{5}} & 0 & \;\;0 & \;\;0 \\
  0 & -e^{u^{6}} & \frac{5}{2}e^{u^{1}} & \frac{5}{3}e^{u^{4}} &\;\; 0& \;\;0 \\
   0 & 0 & 0 & 0 & \;\;0 & \;\;0   \\
    0 & 0 & 0 & 0 & \;\;0 & \;\;0
\end{array} \right), \medskip \\
Z_{8}=\left(\begin{array}{cccccc}
  0 & 0 & 0 & 0 &\;\; 0 & \;\;0 \\
  0 & 0 & \frac{5}{2}e^{u^{1}} & e^{u^{5}} & \;\;0 & \;\;0  \\
  0 & -e^{u^{6}} & 0 & e^{u^{3}}  & \;\;0 & \;\;0 \\
  0 & 0 & 0 & 0 &\;\; 0& \;\;0 \\
   0 & 0 & 0 & 0 & \;\;0 & \;\;0   \\
    0 & 0 & 0 & 0 & \;\;0 & \;\;0
\end{array} \right), &
Z_{9}=\left(\begin{array}{crrrrr}
  0 & 0 &  \;\;0 &\;\; 0 &\;\; 0 & \;\;0 \\
  0 & e^{u^{3}} & e^{u^{5}} & 0 & \;\;0 & \;\;0  \\
  0 & 0 & 0 & 0 & \;\;0 & \;\;0 \\
  0 & 0 & 0 & 0 &\;\; 0& \;\;0 \\
   0 & 0 & 0 & 0 & \;\;0 & \;\;0   \\
    0 & 0 & 0 & 0 & \;\;0 & \;\;0
\end{array} \right),
\\ Z_{10}=\left(\begin{array}{crrrrr}
  0 & \;\;0 & \;\;0 & \;\;0 &\;\; 0 & \;\;0 \\
  0 & \;\;e^{u^{4}} & 0 & 0 & \;\;0 & \;\;0  \\
  0 & \;\;0 & 0 & 0 & \;\;0 & \;\;0 \\
  0 & \;\;0 & 0 & 0 &\;\; 0& \;\;0 \\
   0 & \;\;0 & 0 & 0 & \;\;0 & \;\;0   \\
    0 & \;\;0 & 0 & 0 & \;\;0 & \;\;0
\end{array} \right), &
Z_{11}=\left(\begin{array}{crrrrr}
  0 & 0 &\;\; 0 & \;\;0 &\;\; 0 & \;\;0 \\
  0 & e^{u^{2}} & 0 & 0 & \;\;0 & \;\;0  \\
  0 & 0 & 0 & 0 & \;\;0 & \;\;0 \\
  0 & 0 & 0 & 0 &\;\; 0& \;\;0 \\
   0 & 0 & 0 & 0 & \;\;0 & \;\;0   \\
    0 & 0 & 0 & 0 & \;\;0 & \;\;0
\end{array} \right), \medskip \\  Z_{12}=0.
\end{array}
\end{equation*}

To find the Laplace invariants of Eqs. (\ref{n5}) when $\alpha=7,$
we introduce matrices of size 7.

The matrices $B_{m}, \;\; m=1,2, \ldots ,17\;$ of size 7, are
defined as
\begin{itemize}
\item[--] the first column of the matrices $ B_{m}, \; m\leq 5 $ consist of
arbitrary elements,

\item[--] for $m =6, 7 $ in additional to the first column, the last column
$ B_{m}$
 is also arbitrary,

 \item[--] the
first,  sixth and seventh columns of the matrices $ B_{m}, \; m =
8, 9 $ consist of arbitrary elements,

 \item[--] the
first, second, sixth and seventh columns of the matrices $ B_{m},
\; m = 10, 11 $ consist of arbitrary elements,

\item[--] the first, second, third, sixth and seventh columns of the
matrices $ B_{m}, \; m=12, 13 $ consist of arbitrary elements,

\item[--] for $m \geq 14 $ in additional to the first, second, third, firth,
sixth and seventh  columns, the third column $ B_{m}$
 is also arbitrary,
\end{itemize}

with the other element equal to zero. \vspace{0.5cm}

The matrices  $P_{m}, \;\; m=2,3, \ldots ,12,$ are given by
\begin{equation*}

\end{equation*}

The diagonal matrices $S_{m}$ and $R_{m},$
 are given by \begin{equation*}
\begin{array}{l}
S_{1}=\mbox{diag}\left( \exp(u^{1}), \exp(u^{2}), \exp(u^{3}),
\exp(u^{4}), \exp(u^{5}), \exp(u^{6}), \exp(u^{7}) \right), \medskip \\
S_{2}=\mbox{diag}\left( 0, \exp(u^{1}+u^3), \exp(u^2+u^4),
\exp(u^3+u^4), \exp(u^4+u^5),
 \exp(u^5+u^6), \exp(u^6+u^7)\right), \medskip \\ S_{3}=\mbox{diag}\left( 0, 4\exp(u^{2}+u^3+u^4),  \exp(u^1+u^3+u^4),
\exp(u^2+u^4+u^5), \exp(u^4+u^5+u^6), \right. \medskip \\ \left.
\exp(u^5+u^6+u^7)  \right), \quad S_{4}=\mbox{diag}\left( 0,
9\exp(u^{1}+u^{2}+u^3+u^4), 4\exp(u^2+u^3+u^4+u^{5}), \right.
\medskip \\ \left. \exp(u^1+u^3+u^4+u^{5}),  \exp(u^2+u^4+u^5+u^{6}),  \exp(u^3+u^4+u^5+u^{6}),
\exp(u^4+u^5+u^6+u^{7})
 \right), \medskip \\ S_{5}=\mbox{diag}\left( 0, 4\exp(u^{2}+u^{3}+2u^4+u^5),
9\exp(u^{1}+u^2+u^3+u^4+u^{5}), 4\exp(u^{2}+u^3+u^4+u^5+u^{6}),
\right. \medskip \\ \left. \exp(u^{1}+u^3+u^4+u^5+u^{6}),
\exp(u^{2}+u^4+u^5+u^6+u^{7}),\exp(u^{3}+u^4+u^5+u^6+u^{7})  \right), \medskip \\
S_{6}=\mbox{diag}\left( 0, 25\exp(u^{1}+u^{2}+u^{3}+2u^4+u^5),
4\exp(u^{2}+u^3+2u^4+u^5+u^{6}), \right. \medskip \\
\end{array}
\end{equation*}
\begin{equation*}
\begin{array}{l}
R_{4}=\mbox{diag}\left( 0, u^{1}_{y}+u^{2}_{y}+u^3_{y}+u^4_{y},
u^2_{y}+u^3_{y}+u^4_{y}+u^{5}_{y},
u^1_{y}+u^3_{y}+u^4_{y}+u^{5}_{y},
u^2_{y}+u^4_{y}+u^5_{y}+u^{6}_{y}, \right. \medskip \\ \left.
u^3_{y}+u^4_{y}+u^5_{y}+u^{6}_{y},
u^4_{y}+u^5_{y}+u^6_{y}+u^{7}_{y}
 \right), \medskip \\ R_{5}=\mbox{diag}\left( 0, u^{2}_{y}+u^{3}_{y}+2u^4_{y}+u^5_{y},
u^{1}_{y}+u^2_{y}+u^3_{y}+u^4_{y}+u^{5}_{y},
u^{2}_{y}+u^3_{y}+u^4_{y}+ u^5_{y}+u^{6}_{y}, \right.\medskip \\
\left.  u^{1}_{y}+u^3_{y}+u^4_{y}+u^5_{y}+u^{6}_{y},
u^{2}_{y}+u^4_{y}+u^5_{y}+u^6_{y}+u^{7}_{y},u^{3}_{y}+u^4_{y}+u^5_{y}+u^6_{y}
+u^{7}_{y}  \right), \medskip\\
R_{6}=\mbox{diag}\left( 0,
u^{1}_{y}+u^{2}_{y}+u^{3}_{y}+2u^4_{y}+u^5_{y},
u^{2}_{y}+u^3_{y}+2u^4_{y}+u^5_{y}+u^{6}_{y},
u^{1}_{y}+u^{2}_{y}+u^3_{y}+ \right. \medskip \\ \left. +
u^4_{y}+u^5_{y}+u^{6}_{y},
u^{2}_{y}+u^3_{y}+u^4_{y}+u^5_{y}+u^{6}_{y}+u^{7}_{y}, u^1_{y}+u^{3}_{y}+u^4_{y}+u^5_{y}+u^6_{y}+u^{7}_{y}, 0  \right),\medskip \\
R_{7}=\mbox{diag}\left( 0,
u^{1}_{y}+u^{2}_{y}+2u^{3}_{y}+2u^4_{y}+u^5_{y},
u^{1}_{y}+u^{2}_{y}+u^3_{y}+2u^4_{y}+u^5_{y}+u^{6}_{y},
u^{2}_{y}+u^{3}_{y}+2u^4_{y}+u^5_{y}+u^{6}_{y}+u^7_{y}, \right.
\medskip \\ \left.
u^{1}_{y}+u^{2}_{y}+u^3_{y}+u^4_{y}+u^5_{y}+u^{6}_{y}+u^{7}_{y},
u^2_{y}+u^{3}_{y}+2u^4_{y}+2u^5_{y}+u^6_{y}+u^{7}_{y}, 0  \right), \medskip \\
\end{array}
\end{equation*}
\vspace{0.5cm}

The matrices $Z_{m}, \;\; m=1,2, \ldots ,18$, are given by$
\;\;Z_{1}=U,$
$$
%
 $$

The nonzero elements of the matrices
$D_{m}=\left(d_{ij}\right)_{i,j=1}^{7}, \; m=2,3, \ldots ,18, $
are determined by the formulas
$$
\begin{array}{l}
d^{3}_{2 1}=-e^{u^{3}}; \;\; d^{7}_{5 7}=2e^{u^{2}}-2e^{u^{3}};
\;\; d^{7}_{6 7}=e^{u^{1}}; \;\; d^{9}_{6 2}=\frac{5}{2}e^{u^{1}};
\; \; d^{9}_{6 4}=-\frac{5}{3}e^{u^{4}}; \medskip \\ d^{9}_{6
5}=-e^{u^{7}}; \quad  d^{10}_{5 2}=\frac{1}{2}e^{u^{1}}; \quad
d^{13}_{3 4}=-e^{u^{7}}; \quad d^{15}_{5 4}=e^{u^{2}}.
\end{array}
 $$

 Finally, the matrices $G_{m}, \;\; m=2,3, \ldots ,18,$ are given by $\;\;G_{2}=J_{7};$
 $$
%
 $$

To find the Laplace invariants of Eqs. (\ref{n5}) when $\alpha=8,$
we introduce matrices of size 8.

The matrices $P_{m},\;\; m=2,3, \ldots ,30,$  are given by
$$
%
 \right),
 $$
 $ P_{m}=0 ,\;\;\; m=25,26, \ldots ,30.$
\vspace{0.5cm}

The diagonal matrices $S_{m}$ and $R_{m},$ are given by
  $$
 \begin{array}{l}
 S_{1}=\mbox{diag}\left( \exp(u^{1}), \exp(u^{2}), \exp(u^{3}),
\exp(u^{4}), \exp(u^{5}), \exp(u^{6}), \exp(u^{7}), \exp(u^{8}) \right), \medskip \\
 S_{2}= \mbox{diag}\left( 0,\exp(u^{1}+u^{3}),\exp(u^{2}+u^{4}),\exp(u^{3}+u^{4}),
\exp(u^{4}+u^{5}),\exp(u^{5}+u^{6}),\exp(u^{6}+u^{7}),\right.
\medskip \\ \left. \exp(u^{7}+u^{8})
\right), \quad
S_{3}= \mbox{diag}\left( 0,4\exp(u^{2}+u^{3}+u^{4}),\exp(u^{1}+u^{3}+u^{4}),\exp(u^{2}+u^{4}+u^{5}),\right. \medskip \\
\left.
\exp(u^{3}+u^{4}+u^{5}),\exp(u^{4}+u^{5}+u^{6}),\exp(u^{5}+u^{6}+u^{7}),\exp(u^{6}+u^{7}+u^{8})
\right), \medskip \\
S_{4}= \mbox{diag}\left(
0,9\exp(u^{1}+u^{2}+u^{3}+u^{4}),4\exp(u^{2}+u^{3}+u^{4}+u^{5}),\exp(u^{1}
+u^{3}+u^{4}+u^{5}), \right.
\medskip \\ \left.
\exp(u^{2}+u^{4}+u^{5}+u^{6}),\exp(u^{3}+u^{4}+u^{5}+u^{6}),
\exp(u^{4}+u^{5}+u^{6}+u^{7}),\exp(u^{5}+u^{6}+u^{7}+u^{8})
\right), \\
S_{5}= \mbox{diag}\left(
0,4\exp(u^{2}+u^{3}+2u^{4}+u^{5}),9\exp(u^{1}+u^{2}+u^{3}+u^{4}+u^{5}),
4\exp(u^{2}+u^{3}+u^{4}+u^{5}+u^{6}),\right.
\medskip \\ \left. \exp(u^{1}+u^{3}+u^{4}+u^{5}+u^{6}), \exp(u^{2}+u^{4}+u^{5}+u^{6}+u^{7}),
\exp(u^{3}+u^{4}+u^{5}+u^{6}+u^{7}),\right.
\medskip \\
\end{array}
 $$
 $$
 \begin{array}{l}
  S_{11}= \mbox{diag}\left(
0,49\exp(u^{1}+2u^{2}+2u^{3}+3u^{4}+2u^{5}+u^{6}),
49\exp(u^{1}+u^{2}+2u^{3}+3u^{4}+2u^{5}+u^{6}+u^{7}), \right.
\medskip \\ \left.
4\exp(u^{2}+u^{3}+2u^{4}+2u^{5}+2u^{6}+2u^{7}+u^{8}),
25\exp(u^{1}+u^{2}+2u^{3}+2u^{4}+2u^{5}+u^{6}+u^{7}+u^{8}),
\right.
\medskip \\ \left.
49\exp(u^{1}+u^{2}+u^{3}+2u^{4}+2u^{5}+2u^{6}+u^{7}+u^{8}),
49\exp(u^{1}+u^{2}+2u^{3}+ 2u^{4}+2u^{5}+2u^{6}+u^{7}), 0 \right), \medskip \\
S_{12}= \mbox{diag}\left(
0,49\exp(u^{1}+2u^{2}+2u^{3}+3u^{4}+2u^{5}+u^{6}+u^{7}),
49\exp(u^{1} +u^{2}+2u^{3}+3u^{4}+2u^{5}+\right.
\medskip \\ \left. +2u^{6}+u^{7}),
4\exp(u^{1}+u^{2}+u^{3}+2u^{4}+2u^{5} +2u^{6}+2u^{7}+u^{8}),
49\exp(u^{1}+u^{2}+2u^{3}+3u^{4}+2u^{5}+\right.
\medskip \\ \left.+u^{6}+u^{7}+u^{8}),
 49\exp(u^{1}+u^{2}+2u^{3}+2u^{4}+2u^{5}+2u^{6}+u^{7}+u^{8}),0,0
\right), \medskip \\
S_{13}= \mbox{diag}\left(
0,49\exp(u^{1}+2u^{2}+2u^{3}+3u^{4}+2u^{5}+2u^{6}+u^{7}),
49\exp(u^{1}+u^{2}+2u^{3}+3u^{4}+3u^{5}+\right.
\medskip \\ \left.+2u^{6}+u^{7}),
4\exp(u^{1}+u^{2}+2u^{3}+2u^{4}+2u^{5}+2u^{6}+2u^{7}+u^{8}),
49\exp(u^{1}+2u^{2}+2u^{3}+3u^{4}+\right.
\medskip \\
\end{array}
$$
$$
\begin{array}{l}
S_{23}= \mbox{diag}\left( 0,49\exp(2u^{1}+2u^{2}+4u^{3}+
5u^{4}+4u^{5}+3u^{6}+2u^{7}+u^{8}), 49\exp(2u^{1}+3u^{2}+\right.
\medskip \\ \left.+ 3u^{3}+5u^{4}+4u^{5}+3u^{6}+2u^{7}+u^{8}), 0,0,0,0,0 \right),  \medskip \\
S_{24}= \mbox{diag}\left( 0,0,49\exp(2u^{1}+3u^{2}+4u^{3}+
5u^{4}+4u^{5}+3u^{6}+2u^{7}+u^{8}), 0,0,0,0,0 \right), \medskip\\
S_{25}= \mbox{diag}\left( 0,0,49\exp(2u^{1}+3u^{2}+4u^{3}+
6u^{4}+4u^{5}+3u^{6}+2u^{7}+u^{8}), 0,0,0,0,0 \right),  \medskip \\
S_{26}= \mbox{diag}\left( 0,0,49\exp(2u^{1}+3u^{2}+4u^{3}+
6u^{4}+5u^{5}+3u^{6}+2u^{7}+u^{8}),  0,0,0,0,0 \right),\medskip \\
S_{27}= \mbox{diag}\left( 0,0,49\exp(2u^{1}+3u^{2}+4u^{3}+
6u^{4}+5u^{5}+ 4u^{6}+2u^{7}+u^{8}), 0,0,0,0,0 \right), \medskip
\\ S_{28}= \mbox{diag}\left( 0,0,49\exp(2u^{1}+3u^{2}+
4u^{3}+ 6u^{4}+5u^{5}+4u^{6}+3u^{7}+u^{8}),
 0,0,0,0,0 \right), \medskip \\
S_{29}= \mbox{diag}\left( 0,0,49\exp(2u^{1}+3u^{2}+4u^{3}+
6u^{4}+5u^{5}+4u^{6}+3u^{7}+2u^{8}),  0,0,0,0,0 \right),  \quad S_{30}=0,\\
\end{array}
$$
$$
\begin{array}{l}
R_{11}= \mbox{diag}\left( 0,
u^{1}_{y}+2u^{2}_{y}+2u^{3}_{y}+3u^{4}_{y}+2u^{5}_{y}+u^{6}_{y},
\right. u^{1}_{y}+u^{2}_{y}+2u^{3}_{y}+3u^{4}_{y}+
2u^{5}_{y}+u^{6}_{y}+u^{7}_{y}, \medskip \\
u^{2}_{y}+u^{3}_{y}+2u^{4}_{y}+2u^{5}_{y}+u^{6}_{y}+2u^{7}_{y}+u^{8}_{y},
u^{1}_{y}+u^{2}_{y}+2u^{3}_{y}
+2u^{4}_{y}+2u^{5}_{y}+u^{6}_{y}+u^{7}_{y}+u^{8}_{y}, \medskip \\
u^{1}_{y}+u^{2}_{y}+u^{3}_{y}+2u^{4}_{y}+2u^{5}_{y}+2u^{6}_{y}+u^{7}_{y}+u^{8}_{y},
 u^{1}_{y}+u^{2}_{y}+2u^{3}_{y}+
\left. 2u^{4}_{y}+2u^{5}_{y}+2u^{6}_{y}+u^{7}_{y}, 0 \right), \medskip \\
R_{12}= \mbox{diag}\left( 0,
u^{1}_{y}+2u^{2}_{y}+2u^{3}_{y}+3u^{4}_{y}+2u^{5}_{y}+u^{6}_{y}+u^{7}_{y},
u^{1}_{y}+\right.u^{2}_{y}+2u^{3}_{y} +
3u^{4}_{y}+2u^{5}_{y}+2u^{6}_{y}+u^{7}_{y}, \medskip \\
u^{1}_{y}+u^{2}_{y}+u^{3}_{y}+2u^{4}_{y}+2u^{5}_{y}+2u^{6}_{y}+2u^{7}_{y}+u^{8}_{y},
u^{1}_{y}+u^{2}_{y}+2u^{3}_{y}+3u^{4}_{y}+2u^{5}_{y}+u^{6}_{y}+u^{7}_{y}+u^{8}_{y},
\medskip \\  u^{1}_{y}+u^{2}_{y}+2u^{3}_{y}+2u^{4}_{y}+2u^{5}_{y}+
\left. +2u^{6}_{y}+u^{7}_{y}+u^{8}_{y},0,0 \right), \medskip \\
R_{13}= \mbox{diag}\left( 0,
u^{1}_{y}+2u^{2}_{y}+2u^{3}_{y}+3u^{4}_{y}+2u^{5}_{y}+
2u^{6}_{y}+u^{7}_{y},
u^{1}_{y}+u^{2}_{y}+2u^{3}_{y}+3u^{4}_{y}+3u^{5}_{y}+2u^{6}_{y}+u^{7}_{y},
\right. \medskip \\
u^{1}_{y}+u^{2}_{y}+2u^{3}_{y}+2u^{4}_{y}+2u^{5}_{y}+2u^{6}_{y}+2u^{7}_{y}+u^{8}_{y},
u^{1}_{y}+2u^{2}_{y}+2u^{3}_{y}+3u^{4}_{y}+2u^{5}_{y}+u^{6}_{y}+u^{7}_{y}+u^{8}_{y}, \medskip \\
\end{array}
$$
$$
\begin{array}{l}
 R_{27}= \mbox{diag}\left(
0,0, 2u^{1}_{y}+3u^{2}_{y}+4u^{3}_{y}+ 6u^{4}_{y}+5u^{5}_{y}+
4u^{6}_{y}+2u^{7}_{y}+u^{8}_{y}, 0,0,0,0,0 \right),\medskip \\
R_{28}= \mbox{diag}\left( 0,0, 2u^{1}_{y}+3u^{2}_{y}+4u^{3}_{y}+
6u^{4}_{y}+5u^{5}_{y}+4u^{6}_{y}+3u^{7}_{y}+u^{8}_{y},
0,0,0,0,0 \right),\medskip \\
R_{29}= \mbox{diag}\left( 0,0,2u^{1}_{y}+3u^{2}_{y}+4u^{3}_{y}+
6u^{4}_{y}+5u^{5}_{y}+4u^{6}_{y}+3u^{7}_{y}+2u^{8}_{y}, 0,0,0,0,0
\right).
 \end{array}
$$
\vspace{0.5cm}

The matrices  $Z_{m}, \;\; m=1,2, \ldots ,30,$ are given by $ \;\;
Z_{1}=U,$
$$
Z_{2}=\left(\begin{array}{cccccccc}
  0 & 0 & 0 & 0 & 0 & 0 & 0 & 0 \\
  -e^{u^{3}} & 0 & e^{u^{1}} & 0 & 0 & 0 & 0 & 0 \\
  0 & -e^{u^{4}} & 0 & e^{u^{2}} & 0 & 0 & 0 & 0 \\
  0 & 0 & -e^{u^{4}} & e^{u^{3}} & 0 & 0 & 0 & 0 \\
  0 & 0 & 0 & -e^{u^{5}} & e^{u^{4}} & 0 & 0 & 0 \\
  0 & 0 & 0 & 0 & -e^{u^{6}} & e^{u^{5}} & 0 & 0 \\
  0 & 0 & 0 & 0 & 0 & -e^{u^{7}} & e^{u^{6}} & 0 \\
  0 & 0 & 0 & 0 & 0 & 0 & -e^{u^{8}} & e^{u^{7}}
\end{array} \right),
$$
$$
Z_{3}=\left(\begin{array}{cccccccc}
  0 & 0 & 0 & 0 & 0 & 0 & 0 & 0 \\
  0 & 0 & 2e^{u^{3}} & 2e^{u^{2}} & 0 & 0 & 0 & 0 \\
  0 & -e^{u^{4}} & 0 & e^{u^{1}} & 0 & 0 & 0 & 0 \\
  0 & 0 & -e^{u^{5}} & 0 & e^{u^{2}} & 0 & 0 & 0 \\
  0 & 0 & 0 & -e^{u^{5}} & e^{u^{3}} & 0 & 0 & 0 \\
  0 & 0 & 0 & 0 & -e^{u^{6}} & e^{u^{4}} & 0 & 0 \\
  0 & 0 & 0 & 0 & 0 & -e^{u^{7}} & e^{u^{5}} & 0 \\
  0 & 0 & 0 & 0 & 0 & 0 & -e^{u^{8}} & e^{u^{6}}
\end{array} \right),
$$
$$
Z_{4}=\left(\begin{array}{cccccccc}
  0 & 0 & 0 & 0 & 0 & 0 & 0 & 0 \\
  0 & \frac{3}{2}e^{u^{1}} & 3e^{u^{2}} & 0 & 0 & 0 & 0 & 0 \\
  0 & -e^{u^{5}} & 0 & 2e^{u^{3}} & 2e^{u^{2}} & 0 & 0 & 0 \\
  0 & 0 & -e^{u^{5}} & 0 & e^{u^{1}} & 0 & 0 & 0 \\
  0 & 0 & 0 & -e^{u^{6}} & 0& e^{u^{2}} & 0 & 0 \\
  0 & 0 & 0 & 0 & -e^{u^{6}} & e^{u^{3}} & 0 & 0 \\
  0 & 0 & 0 & 0 & 0 & -e^{u^{7}} & e^{u^{4}} & 0 \\
  0 & 0 & 0 & 0 & 0 & 0 & -e^{u^{8}} & e^{u^{5}}
\end{array} \right),
$$
$$
Z_{5}=\left(\begin{array}{cccccccc}
  0 & 0 & 0 & 0 & 0 & 0 & 0 & 0 \\
 0 & 0 & e^{u^{4}} & 0 & 0 & 0 & 0 & 0 \\
  0 & -e^{u^{5}} & \frac{3}{2}e^{u^{1}} & 3e^{u^{2}}  & 0 & 0 & 0 & 0 \\
  0 & 0 & -e^{u^{6}} & 0 & 2e^{u^{3}} & 2e^{u^{2}}  & 0 & 0 \\
  0 & 0 &  0& -e^{u^{6}} & 0 & e^{u^{1}}  & 0 & 0 \\
  0 & 0 & 0 & 0& -e^{u^{7}} & 0& e^{u^{2}} & 0 \\
   0 & 0 & 0 & 0 & 0 & -e^{u^{7}} & e^{u^{3}} & 0 \\
  0 & 0 & 0 & 0 & 0 & 0 & -e^{u^{8}} & e^{u^{4}}
\end{array} \right),
$$
$$
Z_{6}=\left(\begin{array}{lccccccc}
  0\;\; & 0 & 0 & 0 & 0 & 0 & 0 & 0 \\
 0 & \frac{5}{2}e^{u^{1}} & \frac{5}{3}e^{u^{4}} & 0 & 0 & 0 & 0 & 0
 \\
 0 & -e^{u^{6}} & 0 & e^{u^{4}} & 0 & 0 & 0  & 0 \\
  0 & 0 & -e^{u^{6}} & \frac{3}{2}e^{u^{1}} & 3e^{u^{2}} & 0 & 0 & 0 \\
  0 & 0 & 0 & -e^{u^{7}} & 0 & 2e^{u^{3}} & 2e^{u^{2}}  & 0 \\
  0 & 0 &  0 & 0& -e^{u^{7}} & 0 & e^{u^{1}}  & 0 \\
  0 & 0 & 0 & 0 & 0& -e^{u^{8}} & 0& e^{u^{2}}\\
     0 & 0 & 0 & 0 & 0 & 0 & -e^{u^{8}} & e^{u^{3}}
\end{array} \right),
$$
$$
Z_{7}=\left(\begin{array}{lccccccc}
  0\;\; & 0 & 0 & 0 & 0 & 0 & 0 & 0 \\
   0\;\; & e^{u^{3}} & 0 & 0 & 0 & 0 & 0 & 0 \\
    0\;\; & 0 & e^{u^{5}} & 0 & 0 & 0 & 0 & 0 \\
 0 & -e^{u^{6}} & \frac{5}{2}e^{u^{1}} & \frac{5}{3}e^{u^{4}} & 0 & 0 & 0 & 0
 \\
 0 & 0 & -e^{u^{7}} & 0 & e^{u^{4}} & 0  & 0  & 0 \\
  0 & 0 & 0 & -e^{u^{7}} & \frac{3}{2}e^{u^{1}} & 3e^{u^{2}} & 0 & 0 \\
  0 & 0 & 0 & 0&  -e^{u^{8}} & 0 & 2e^{u^{3}} & 2e^{u^{2}} \\
      0 & 0 & 0 & 0 & 0  & -e^{u^{8}} & 0 & e^{u^{1}}
\end{array} \right),
$$
$$
Z_{8}=\left(\begin{array}{lccccccc}
  0\;\; & 0 & 0 & 0 & 0 & 0 & 0 & 0 \\
   0 & 0 & \frac{7}{2}e^{u^{1}} & \frac{7}{5}e^{u^{5}} & 0 & 0 & 0 &
   0\\
   0\;\; & -e^{u^{6}} & 0 & e^{u^{3}} & 0 & 0 & 0 & 0 \\
    0\;\; & 0 & -e^{u^{7}} & 0 & e^{u^{5}} & 0 & 0 & 0 \\
 0 & 0 & 0 & -e^{u^{7}} & \frac{5}{2}e^{u^{1}} & \frac{5}{3}e^{u^{4}} & 0 &  0
 \\
 0 & 0 & 0 & 0 & -e^{u^{8}} & 0 & e^{u^{4}} & 0   \\
  0 & 0 & 0 & 0 & 0 & -e^{u^{8}} & \frac{3}{2}e^{u^{1}} &
  3e^{u^{2}}\\
  0\;\; & 0 & 0 & 0 & 0 & 0 & 0 & 0
  \end{array} \right),
$$
$$
Z_{9}=\left(\begin{array}{lccccccc}
  0\;\; & 0 & 0 & 0 & 0 & 0 & 0 & 0 \\
0\;\; & 0 & 0 & e^{u^{6}} & 0 & 0 & 0 & 0 \\
   0 & -e^{u^{7}} & 0 & \frac{7}{2}e^{u^{1}} & \frac{7}{5}e^{u^{5}} & 0 & 0 & 0 \\
      0\;\; & 0 & -e^{u^{7}} & 0 & e^{u^{3}} & 0 & 0 & 0 \\
 0\;\; & 0 & 0 &  -e^{u^{8}} & 0 & e^{u^{5}} & 0 & 0  \\
 0 & 0 & 0 & 0 & -e^{u^{8}} & \frac{5}{2}e^{u^{1}} & \frac{5}{3}e^{u^{4}} & 0
 \\
 0 &  e^{u^{3}}& \frac{7}{5}e^{u^{5}} & 0 & 0& 0 & 0 & 0   \\
   0\;\; & 0 & 0 & 0 & 0 & 0 & 0 & 0
  \end{array} \right),
$$
$$
Z_{10}=\left(\begin{array}{lccccccr}
  0\;\; & 0 & 0 & 0 & 0 & 0 & 0 & \;\; 0 \\
0\;\; & 0 & 0 & 0 & 0 & 0 & e^{u^{4}} & 0 \\
0\;\; & -e^{u^{8}} & 0  & 0 & e^{u^{6}}  & 0 & 0 & 0 \\
   0 & 0 & -e^{u^{8}} & 0 & \frac{7}{2}e^{u^{1}} & \frac{7}{5}e^{u^{5}} & 0 & 0\\
      0\;\; & 0 & 0 & -e^{u^{8}} & 0 & e^{u^{3}} & 0 & 0  \\
 0\;\; & 0 & -e^{u^{3}}& -\frac{7}{5}e^{u^{5}} & 0 & 0 & e^{u^{7}} & 0  \\
  0 &  \frac{7}{2}e^{u^{1}}& e^{u^{6}} & 0 & 0& 0 & 0 & 0   \\
   0\;\; & 0 & 0 & 0 & 0 & 0 & 0 & 0
  \end{array} \right),
$$
$$
Z_{11}=\left(\begin{array}{lccccccr}
  0\;\; & 0 & 0 & 0 & 0 & 0 & 0 & \;\; 0 \\
0\;\; & e^{u^{2}} & 0 & 0 & 0 & 0 & 0 & 0 \\
0\;\; & e^{u^{7}} & 0  & 0  & 0 & e^{u^{4}}  & 0 & 0  \\
   0 & 0 & -e^{u^{7}} & 0 &0 & 0 & 0 & 0\\
      0\;\; & 0 & 0 & \frac{5}{7}e^{u^{3}} & e^{u^{5}} & \frac{5}{7}e^{u^{8}} & 0 & 0  \\
 0\;\; & 0 & -\frac{7}{2}e^{u^{1}}& -e^{u^{6}} & 0 & 0 & e^{u^{8}} & 0  \\
  0 &  0& 0 & 0 & 0& -e^{u^{6}} & e^{u^{3}}& 0   \\
   0\;\; & 0 & 0 & 0 & 0 & 0 & 0 & 0
  \end{array} \right),
$$
$$
Z_{12}=\left(\begin{array}{lccccccr}
  0\;\; & 0 & 0 & 0 & 0 & 0 & 0 & \;\; 0 \\
0\;\; & e^{u^{7}} & e^{u^{2}} & 0 & 0 & 0 & 0 & 0 \\
0\;\; & 0 & e^{u^{6}} & 0  & 0  & 0 & -e^{u^{4}}  & 0   \\
   0 & 0 & 0 & e^{u^{1}} & 0 & \frac{2}{7}e^{u^{7}} & 0 & 0 \\
      0\;\; & 0 & e^{u^{8}}& 0 & \frac{7}{5}e^{u^{4}} & 0  & 0 & 0  \\
 0\;\; & 0 & 0 & 0 &-\frac{7}{5}e^{u^{6}}& e^{u^{3}} & e^{u^{8}} & 0  \\
0\;\; & 0 & 0 & 0 & 0 & 0 & 0 & 0 \\
   0\;\; & 0 & 0 & 0 & 0 & 0 & 0 & 0
  \end{array} \right),
$$
$$
Z_{13}=\left(\begin{array}{lccccccr}
  0\;\; & 0 & 0 & 0 & 0 & 0 & 0 & \;\; 0 \\
0\;\; & e^{u^{6}} & e^{u^{2}} & 0 & 0 & 0 & 0 & 0 \\
0\;\; & 0 & e^{u^{5}} & 0  & 0  & 0 & 0  & 0   \\
   0 & 0 & 0 & e^{u^{3}} & 0 & \frac{2}{7}e^{u^{7}} & 0 & 0 \\
      0\;\; &  e^{u^{8}} & 0& 0 & e^{u^{2}} & 0  & 0 & 0  \\
 0\;\; & 0 & - e^{u^{8}} & 0 & -e^{u^{6}}& e^{u^{4}} &0 & 0  \\
0\;\; & 0 & 0 & 0 & 0 & 0 & 0 & 0 \\
   0\;\; & 0 & 0 & 0 & 0 & 0 & 0 & 0
  \end{array} \right),
$$
$$
Z_{14}=\left(\begin{array}{lccccccr}
  0\;\; & 0 & 0 & 0 & 0 & 0 & 0 & \;\; 0 \\
0\;\; & e^{u^{5}} & e^{u^{2}} & 0 & 0 & 0 & 0 & 0 \\
0\;\; & 0 & e^{u^{8}} & 0  & 0  & -e^{u^{5}} & 0  & 0   \\
   0 & 0 & 0 & e^{u^{4}} & 0 & \frac{2}{7}e^{u^{7}} & 0 & 0 \\
      0\;\; &  e^{u^{8}} & 0& 0 & e^{u^{6}} & -e^{u^{2}}  & 0 & 0  \\
 0\;\; & 0 & 0 & 0 & 0& 0 &0 & 0  \\
0\;\; & 0 & 0 & 0 & 0 & 0 & 0 & 0 \\
   0\;\; & 0 & 0 & 0 & 0 & 0 & 0 & 0
  \end{array} \right),
$$
$$
Z_{15}=\left(\begin{array}{lccccrrr}
  0\;\; & 0 & 0 & 0 & 0 & \;\;0 & \;\; 0 & \;\; 0 \\
0\;\; & e^{u^{4}} & 0 & 0 & 0 & 0 & 0 & 0 \\
0\;\; & e^{u^{8}} & e^{u^{2}} & 0  & e^{u^{5}}  & 0 & 0  & 0   \\
   0 & 0 & -\frac{2}{7}e^{u^{7}} & e^{u^{5}} & 0 & 0 & 0 & 0 \\
      0\;\; &  0 & 0&  -e^{u^{2}} & \frac{2}{7}e^{u^{7}} & 0& 0 & 0  \\
 0\;\; & 0 & 0 & 0 & 0& 0 &0 & 0  \\
0\;\; & 0 & 0 & 0 & 0 & 0 & 0 & 0 \\
   0\;\; & 0 & 0 & 0 & 0 & 0 & 0 & 0
  \end{array} \right),
$$
$$
Z_{16}=\left(\begin{array}{lccccrrr}
  0\;\; & 0 & 0 & 0 & 0 & \;\;0 & \;\; 0 & \;\; 0 \\
0\;\; & e^{u^{3}} & 0 & 0 & 0 & 0 & 0 & 0 \\
0\;\; & e^{u^{8}} & e^{u^{4}} & 0  & 0  & 0 & 0  & 0   \\
0\;\; & 0 & 0 & e^{u^{6}} & 0 & 0 & 0 & 0 \\
   0 & 0 & \frac{2}{7}e^{u^{7}} &- e^{u^{2}} & e^{u^{5}} & 0 & 0 & 0 \\
    0\;\; & 0 & 0 & 0 & 0& 0 &0 & 0  \\
0\;\; & 0 & 0 & 0 & 0 & 0 & 0 & 0 \\
   0\;\; & 0 & 0 & 0 & 0 & 0 & 0 & 0
  \end{array} \right),
$$
$$
Z_{17}=\left(\begin{array}{lccccrrr}
  0\;\; & 0 & 0 & 0 & 0 & \;\;0 & \;\; 0 & \;\; 0 \\
0\;\; & e^{u^{1}} & 0 & 0 & 0 & 0 & 0 & 0 \\
0\;\; & e^{u^{8}} & e^{u^{3}} & 0  & 0  & 0 & 0  & 0   \\
0\;\; & 0 & 0 & -e^{u^{2}} & e^{u^{6}} & 0 & 0 & 0 \\
   0 & 0 & \frac{2}{7}e^{u^{7}} &0 & e^{u^{4}} & 0 & 0 & 0 \\
    0\;\; & 0 & 0 & 0 & 0& 0 &0 & 0  \\
0\;\; & 0 & 0 & 0 & 0 & 0 & 0 & 0 \\
   0\;\; & 0 & 0 & 0 & 0 & 0 & 0 & 0
  \end{array} \right),
$$
$$
Z_{18}=\left(\begin{array}{lccccrrr}
  0\;\; & 0 & 0 & 0 & 0 & \;\;0 & \;\; 0 & \;\; 0 \\
0\;\; & e^{u^{8}} &  e^{u^{1}} & 0 & 0 & 0 & 0 & 0 \\
0\;\; & 0 & e^{u^{7}} & 0  & \frac{7}{2} e^{u^{3}}  & 0 & 0  & 0   \\
0\;\; & 0 & 0 & e^{u^{4}} & e^{u^{6}} & 0 & 0 & 0 \\
   0\;\; & 0 & 0 & 0 & 0& 0 &0 & 0  \\
    0\;\; & 0 & 0 & 0 & 0& 0 &0 & 0  \\
0\;\; & 0 & 0 & 0 & 0 & 0 & 0 & 0 \\
   0\;\; & 0 & 0 & 0 & 0 & 0 & 0 & 0
  \end{array} \right),
$$
$$
Z_{19}=\left(\begin{array}{lccccrrr}
  0\;\; & 0 & 0 & 0 & 0 & \;\;0 & \;\; 0 & \;\; 0 \\
0\;\; & e^{u^{7}} &  e^{u^{1}} & 0 & 0 & 0 & 0 & 0 \\
0\;\; & 0 & e^{u^{6}}   & \frac{7}{2} e^{u^{3}} & 0 & 0 & 0  & 0   \\
0\;\; & 0 & 0 & e^{u^{5}} & 0 & 0 & 0 & 0 \\
   0\;\; & 0 & 0 & 0 & 0& 0 &0 & 0  \\
    0\;\; & 0 & 0 & 0 & 0& 0 &0 & 0  \\
0\;\; & 0 & 0 & 0 & 0 & 0 & 0 & 0 \\
   0\;\; & 0 & 0 & 0 & 0 & 0 & 0 & 0
  \end{array} \right),
$$
$$
Z_{20}=\left(\begin{array}{lccccrrr}
  0\;\; & 0 & 0 & 0 & 0 & \;\;0 & \;\; 0 & \;\; 0 \\
0\;\; & e^{u^{6}} &  e^{u^{1}} & 0 & 0 & 0 & 0 & 0 \\
0\;\; & 0 & e^{u^{5}}   & \frac{7}{2} e^{u^{3}} & 0 & 0 & 0  & 0   \\
0\;\; & 0 & 0 & 0 & 0 & 0 & 0 & 0 \\
   0\;\; & 0 & 0 & 0 & 0& 0 &0 & 0  \\
    0\;\; & 0 & 0 & 0 & 0& 0 &0 & 0  \\
0\;\; & 0 & 0 & 0 & 0 & 0 & 0 & 0 \\
   0\;\; & 0 & 0 & 0 & 0 & 0 & 0 & 0
  \end{array} \right),
$$
$$ Z_{21}=\left(\begin{array}{lccrrrrr}
  0\;\; & 0 & 0 & \;\;0 & \;\;0 & \;\;0 & \;\; 0 & \;\; 0 \\
0\;\; & e^{u^{5}} &  e^{u^{1}} & 0 & 0 & 0 & 0 & 0 \\
0\;\; & 0 & e^{u^{4}}   & 0 & 0 & 0 & 0  & 0   \\
0\;\; & 0 & 0 & 0 & 0 & 0 & 0 & 0 \\
   0\;\; & 0 & 0 & 0 & 0& 0 &0 & 0  \\
    0\;\; & 0 & 0 & 0 & 0& 0 &0 & 0  \\
0\;\; & 0 & 0 & 0 & 0 & 0 & 0 & 0 \\
   0\;\; & 0 & 0 & 0 & 0 & 0 & 0 & 0
  \end{array} \right),
$$
$$ Z_{22}=\left(\begin{array}{lccrrrrr}
  0\;\; & 0 & 0 & \;\;0 & \;\;0 & \;\;0 & \;\; 0 & \;\; 0 \\
0\;\; & e^{u^{4}} &  e^{u^{1}} & 0 & 0 & 0 & 0 & 0 \\
0\;\; & 0 & e^{u^{2}}   & 0 & 0 & 0 & 0  & 0   \\
0\;\; & 0 & 0 & 0 & 0 & 0 & 0 & 0 \\
   0\;\; & 0 & 0 & 0 & 0& 0 &0 & 0  \\
    0\;\; & 0 & 0 & 0 & 0& 0 &0 & 0  \\
0\;\; & 0 & 0 & 0 & 0 & 0 & 0 & 0 \\
   0\;\; & 0 & 0 & 0 & 0 & 0 & 0 & 0
  \end{array} \right),
$$
$$ Z_{23}=\left(\begin{array}{lccrrrrr}
  0\;\; & 0 & 0 & \;\;0 & \;\;0 & \;\;0 & \;\; 0 & \;\; 0 \\
0\;\; & e^{u^{3}} &  0 & 0 & 0 & 0 & 0 & 0 \\
0\;\; & e^{u^{2}} & e^{u^{1}}   & 0 & 0 & 0 & 0  & 0   \\
0\;\; & 0 & 0 & 0 & 0 & 0 & 0 & 0 \\
   0\;\; & 0 & 0 & 0 & 0& 0 &0 & 0  \\
    0\;\; & 0 & 0 & 0 & 0& 0 &0 & 0  \\
0\;\; & 0 & 0 & 0 & 0 & 0 & 0 & 0 \\
   0\;\; & 0 & 0 & 0 & 0 & 0 & 0 & 0
  \end{array} \right),
$$
$$ Z_{24}=\left(\begin{array}{lccrrrrr}
  0\;\; & 0 & 0 & \;\;0 & \;\;0 & \;\;0 & \;\; 0 & \;\; 0 \\
0\;\; &0 &  0 & 0 & 0 & 0 & 0 & 0 \\
0\;\; & e^{u^{2}} & e^{u^{3}}   & 0 & 0 & 0 & 0  & 0   \\
0\;\; & 0 & 0 & 0 & 0 & 0 & 0 & 0 \\
   0\;\; & 0 & 0 & 0 & 0& 0 &0 & 0  \\
    0\;\; & 0 & 0 & 0 & 0& 0 &0 & 0  \\
0\;\; & 0 & 0 & 0 & 0 & 0 & 0 & 0 \\
   0\;\; & 0 & 0 & 0 & 0 & 0 & 0 & 0
  \end{array} \right),
$$
$$ Z_{m}=\left(\begin{array}{lrcrrrrr}
  0\;\; & \;\; 0 & 0 & \;\;0 & \;\;0 & \;\;0 & \;\; 0 & \;\; 0 \\
0\;\; &0 &  0 & 0 & 0 & 0 & 0 & 0 \\
0\;\; & 0 & e^{u^{m-21}}   & 0 & 0 & 0 & 0  & 0   \\
0\;\; & 0 & 0 & 0 & 0 & 0 & 0 & 0 \\
   0\;\; & 0 & 0 & 0 & 0& 0 &0 & 0  \\
    0\;\; & 0 & 0 & 0 & 0& 0 &0 & 0  \\
0\;\; & 0 & 0 & 0 & 0 & 0 & 0 & 0 \\
   0\;\; & 0 & 0 & 0 & 0 & 0 & 0 & 0
  \end{array} \right), $$ $ m=25,26,27,28,29,
 \quad Z_{30}=0.
$

\newpage
The nonzero elements of the matrices
$D_{m}=\left(d_{ij}\right)_{i,j=1}^{8}, \; m=2,3, \ldots ,30, $
 are determined by the formulas
$$
\begin{array}{l}
d^{3}_{2 1}=-e^{u^{3}}; \;\; d^{9}_{6 8}=-e^{u^{4}}; \;\; d^{9}_{7
8}=3e^{u^{2}}-\frac{3}{2}e^{u^{1}}; \;\; d^{13}_{3 7 }=-e^{u^{4}};
\; \; d^{13}_{6 7}=e^{u^{8}}; \medskip \\ d^{15}_{3 6}=-e^{u^{5}};
\quad d^{15}_{4 6}=\frac{2}{7}e^{u^{7}}; \quad d^{15}_{5
6}=-e^{u^{2}}; \quad d^{19}_{3 5}=\frac{7}{2}e^{u^{3}}, \quad
d^{19}_{4 5}=e^{u^{6}}; \medskip \\ d^{21}_{3
4}=\frac{7}{2}e^{u^{3}}; \quad d^{25}_{3 2}=e^{u^{2}}.
\end{array}
 $$

The matrices $G_{m}, \;\; m=2,3, \ldots ,30,$ are given by
\vspace{0.3cm}

 $\;\;G_{2}=J_{8};\quad $
 $
 G_{m}=\left(\begin{array}{cc}
   1 & 0 \\
   0 &  J_{7}
 \end{array} \right), \quad  m=3,4,\ldots ,8;$
\vspace{0.3cm}

 $
  G_{m}=\left(\begin{array}{ccc}
   1 & 0   & 0\\
   0 &  J_{6} & 0 \\
   0 & 0 & 1
 \end{array} \right) $  for  $\;\;m=9,10,11; \quad
   G_{m}=E_{8}, \quad m \geq 12.
 $
\vspace{0.3cm}

 The matrices $B_{m}, \;\; m=1,2, \ldots ,29\;$ of
size 8, are defined as
\begin{itemize}
  \item [--] the first column of the matrices $ B_{m}, \; m\leq 7 $ consist of
arbitrary elements,

 \item [--] for $m =8, 9, 10$  and $ 11 $ in additional to the first column, the
last column $ B_{m}$
 is also arbitrary,

 \item [--]  the
first,  seventh and eight columns of the matrices $ B_{m}, \; m =
12, 13 $ consist of arbitrary elements,

  \item [--] the
first, sixth, seventh and eight columns of the matrices $ B_{m},
\; m = 14, 15, 16, 17 $ consist of arbitrary elements,

 \item [--] the first, firth, sixth, seventh and eight columns of the matrices
$ B_{m}, \; m = 18, 19 $ consist of arbitrary elements,

 \item [--] the first, fourth, firth, sixth, seventh and eight columns of the
matrices $ B_{m}, \; m = 20, 21, 22, 23$ consist of arbitrary
elements,

 \item [--] for $m \geq 24 $ in additional to the  first, fourth, firth, sixth,
seventh and eight columns, the second column $ B_{m}$
 is also arbitrary,
\end{itemize}

with the other element equal to zero.

\vspace{0.3cm}

We note that the values of the index $m$ at which the rank of the
generalized invariants
  $X_{m}$ drops
\vspace{0.3cm}

 \noindent coincide with the exponents $ \left
\{\begin{array}{l}
 1, 4, 5, 7, 8, 11 \\
  1, 5, 7, 9, 10, 13, 17 \\
  1, 7, 11, 13, 17, 19, 23, 29
\end{array}  \right.\quad $
systems (\ref{n4}) for $\quad \left \{ \begin{array}{c}
  \alpha=6 \\
  \alpha=7 \\
  \alpha=8
\end{array}, \right.$
\vspace{0.3cm}

 \noindent
 corresponding,  and the number   $\left \{ \begin{array}{c}
  r=12 \\
  r=18 \\
  r=30
\end{array}, \right.$
  for which $X_{m}=0$ is
equal to the Coxeter number [4]. \vspace{1cm}

{\bf Acknowledgments.} The authors are grateful to V.V. Sokolov
for the useful discussions of the results in this work.

This work was supported in part by Russian Foundation for Basic
Research \linebreak (Grant \No~05 -- 01 -- 00775 -- a).
\newpage

\begin{flushleft}
{\bf References}
\end{flushleft}

\begin{enumerate}

\item[{[1]}] V.E. Adler and  S.Ya. Startsev, \emph{Theor.
Math. Phys.}, 121, 1484--1495 (1999).

\item[{[2]}] J.M. Anderson and M. Juras, \emph{Duke.
Math. J.}, 89, 351--375 (1997).

\item[{[3]}] J.M. Anderson and N. Kamran, \emph{Duke.
Math. J.}, 87, 265--319 (1997).

\item[{[4]}] N. Bourbaki, \emph{Groupes et alg\'ebres de Lie} (Chap. 1: Alg\'ebres de Lie, Chap. 2:
Alg\'ebres de Lie libres, Chap. 3: Groupes de Lie; Actualit\'es
scientifiques et industrielles 1285, 1349; \'El\'ements de
math\'ematique fascicules 26, 36), Hermann, Paris (1971, 1972).

\item[{[5]}] A.M. Guryeva and  A.V. Zhiber, \emph{Theor.
Math. Phys.}, 138, 401--421 (2004).

\item[{[6]}] E.V. Ferapontov, \emph{Theor. Math. Phys.}, 110, 68--77
(1997).

\item[{[7]}] A.N. Leznov and M.V. Saveliev, \emph{Group Theory Methods of Integration of Nonlinear Dynamical Systems}
 [in Russian], Nauka, Moscow (1985).

\item[{[8]}] A.B. Shabat and R.I. Yamilov, "Exponentional systems of type II and Cartan
matrices",\linebreak  Preprint \No.~1, Bashkir Branch, USSR Acad.
Sci., Ufa (1981).

\item[{[9]}] A.B. Shabat, \emph{Phys.
Lett. A.}, 200, 121--133 (1995).

\item[{[10]}] V.V. Sokolov and A.V. Zhiber, \emph{Phys.
Lett. A.}, 208, 303--308 (1995).

\item[{[11]}] S.Ya. Startsev, \emph{Theor.
Math. Phys.}, 127, 460--470 (2001).

\item[{[12]}] S.P. Tsarev, \emph{Theor.
Math. Phys.}, 122, 121--133 (2000).

\item[{[13]}] A.V. Zhiber, V.V. Sokolov, and S.Ya. Startsev, \emph{Dokl.
Math.}, 52, 128--130 (1995).

\item[{[14]}] A.V. Zhiber and V.V. Sokolov, \emph{Usp.
Mat. Nauk}, 56, 61--101 (2001).

\end{enumerate}

\end{document}